\documentclass[12pt]{article}
\usepackage{amsfonts}
\usepackage{amssymb}
\usepackage{graphics,amsmath}

\def\hybrid{\topmargin -20pt    \oddsidemargin 0pt
        \headheight 0pt \headsep 0pt
        \textwidth 6.35in       % BS paper
        \textheight 9.25in       % BS paper
        \marginparwidth .875in
        \parskip 5pt plus 1pt   \jot = 1.5ex}

\hybrid

\def\baselinestretch{1.2}

\catcode`\@=11

\def\marginnote#1{}
\newcount\hour
\newcount\minute
\newtoks\amorpm
\hour=\time\divide\hour by60
\minute=\time{\multiply\hour by60 \global\advance\minute by-\hour}
\edef\standardtime{{\ifnum\hour<12 \global\amorpm={am}%
        \else\global\amorpm={pm}\advance\hour by-12 \fi
        \ifnum\hour=0 \hour=12 \fi
        \number\hour:\ifnum\minute<10 0\fi\number\minute\the\amorpm}}
\edef\militarytime{\number\hour:\ifnum\minute<10 0\fi\number\minute}
\def\draftlabel#1{{\@bsphack\if@filesw {\let\thepage\relax
   \xdef\@gtempa{\write\@auxout{\string
      \newlabel{#1}{{\@currentlabel}{\thepage}}}}}\@gtempa
   \if@nobreak \ifvmode\nobreak\fi\fi\fi\@esphack}
        \gdef\@eqnlabel{#1}}
\def\@eqnlabel{}
\def\@vacuum{}
\def\draftmarginnote#1{\marginpar{\raggedright\scriptsize\tt#1}}

\def\draft{\oddsidemargin -.5truein
        \def\@oddfoot{\sl preliminary draft \hfil
        \rm\thepage\hfil\sl\today\quad\militarytime}
        \let\@evenfoot\@oddfoot \overfullrule 3pt
        \let\label=\draftlabel
        \let\marginnote=\draftmarginnote
   \def\@eqnnum{(\theequation)\rlap{\kern\marginparsep\tt\@eqnlabel}%
\global\let\@eqnlabel\@vacuum}  }

\def\preprint{\twocolumn\sloppy\flushbottom\parindent 2em
        \leftmargini 2em\leftmarginv .5em\leftmarginvi .5em
        \oddsidemargin -.5in    \evensidemargin -.5in
        \columnsep .4in \footheight 0pt
        \textwidth 10.in        \topmargin  -.4in
        \headheight 12pt \topskip .4in
        \textheight 6.9in \footskip 0pt
        \def\@oddhead{\thepage\hfil\addtocounter{page}{1}\thepage}
        \let\@evenhead\@oddhead \def\@oddfoot{} \def\@evenfoot{} }

\def\numberbysection{\@addtoreset{equation}{section}
        \def\theequation{\thesection.\arabic{equation}}}

\def\underline#1{\relax\ifmmode\@@underline#1\else
        $\@@underline{\hbox{#1}}$\relax\fi}

\def\titlepage{\@restonecolfalse\if@twocolumn\@restonecoltrue\onecolumn
     \else \newpage \fi \thispagestyle{empty}\c@page\z@
        \def\thefootnote{\fnsymbol{footnote}} }

\def\endtitlepage{\if@restonecol\twocolumn \else \newpage \fi
        \def\thefootnote{\arabic{footnote}}
        \setcounter{footnote}{0}}  %\c@footnote\z@ }

\catcode`@=12
\relax

\def\figcap{\section*{Figure Captions\markboth
        {FIGURECAPTIONS}{FIGURECAPTIONS}}\list
        {Figure \arabic{enumi}:\hfill}{\settowidth\labelwidth{Figure
999:}
        \leftmargin\labelwidth
        \advance\leftmargin\labelsep\usecounter{enumi}}}
 \relax
\def\tablecap{\section*{Table Captions\markboth
        {TABLECAPTIONS}{TABLECAPTIONS}}\list
        {Table \arabic{enumi}:\hfill}{\settowidth\labelwidth{Table
999:}
        \leftmargin\labelwidth
        \advance\leftmargin\labelsep\usecounter{enumi}}}
 \relax
\def\reflist{\section*{References\markboth
        {REFLIST}{REFLIST}}\list
        {[\arabic{enumi}]\hfill}{\settowidth\labelwidth{[999]}
        \leftmargin\labelwidth
        \advance\leftmargin\labelsep\usecounter{enumi}}}
 \relax
\makeatletter
\newcounter{pubctr}
\def\publist{\@ifnextchar[{\@publist}{\@@publist}}
\def\@publist[#1]{\list
        {[\arabic{pubctr}]\hfill}{\settowidth\labelwidth{[999]}
        \leftmargin\labelwidth
        \advance\leftmargin\labelsep
        \@nmbrlisttrue\def\@listctr{pubctr}
        \setcounter{pubctr}{#1}\addtocounter{pubctr}{-1}}}
\def\@@publist{\list
        {[\arabic{pubctr}]\hfill}{\settowidth\labelwidth{[999]}
        \leftmargin\labelwidth
        \advance\leftmargin\labelsep
        \@nmbrlisttrue\def\@listctr{pubctr}}}
 \relax
\makeatother
\newskip\humongous \humongous=0pt plus 1000pt minus 1000pt

\newif\ifdtup

\relax

\def\be{\begin{equation}}
\def\ee{\end{equation}}
\def\ba{\begin{eqnarray}}
\def\ea{\end{eqnarray}}

\def\k{\kappa}

\def\a{\alpha}

\def\b{\beta}

\def\g{\gamma}

\def\d{\delta}

\def\p{\pi}

\def\m{\mu}

\def\l{\lambda}

\def\s{\sigma}

\def\no{\noindent}

\def\IR{\relax{\rm I\kern-.18em R}}
\def\II{\relax{\rm 1\kern-.35em1}}

\renewcommand{\theequation}{\thesection.\arabic{equation}}
\csname @addtoreset\endcsname{equation}{section}

\def\IR{\relax{\rm I\kern-.18em R}}
\def\inv{^{\raise.15ex\hbox{${\scriptscriptstyle -}$}\kern-.05em 1}}

\newcommand{\dpartial}[3][]{\frac{\partial^{#1} #2}{\partial #3 ^{#1}}}

\newcommand{\order}[1]{\mathcal{O} \left( #1 \right)}

\newcommand{\ket}[1]{\left| #1 \right\rangle }
\newcommand{\bra}[1]{\left\langle #1 \right| }
\newcommand{\braket}[2]{\left\langle #1 | #2 \right\rangle }

\newcommand{\granesperado}[3]{\left\langle #1 \left| #2 \right| #3 \right\rangle}

\begin{document}

%\draft

\begin{titlepage}
\begin{center}

\vskip .5in

{\LARGE On correlation functions in the coordinate and the algebraic Bethe ansatz}
\vskip 0.4in

{\bf Rafael Hern\'andez}\phantom{x} and\phantom{x}
 {\bf Juan Miguel Nieto}\phantom{x}
\vskip 0.1in

Departamento de F\'{\i}sica Te\'orica I\\
Universidad Complutense de Madrid\\
$28040$ Madrid, Spain\\

{\footnotesize{\tt rafael.hernandez@fis.ucm.es, juanieto@ucm.es}}

\end{center}

\vskip .4in

\centerline{\bf Abstract}
\vskip .1in
\no
The Bethe ansatz, both in its coordinate and its algebraic version, is an exceptional method to compute the eigenvectors and eigenvalues of integrable systems. 
However, computing correlation functions using the eigenvectors thus constructed complicates rather fast. In this article, we will compute some simple correlation functions 
for the isotropic Heisenberg spin chain to highlight the shortcomings of both Bethe ans\"atze. In order to compare the results obtained from each approach, a discussion 
on the normalization of states in each ansatz will be required. We will show that the analysis can be extended to the long-range spin chain governing the spectrum 
of anomalous dimensions of single trace operators in four-dimensional Yang-Mills with maximal supersymmetry. 

\noindent

\vskip .4in
\noindent

\end{titlepage}
\vfill
\eject

\def\baselinestretch{1.2}

%%%%%%%%%%%%%%%%%%%%%%%%%%%%%%%%%%%%%%%%%%%%%%%%%%%

\baselineskip 20pt

%%%%%%%%%%%%%%%%%%%%%%%%%%%%%%%%%%%%%%%%%%%%%%%%%%%%%%%%%%%%%%%%%%%%%%%%
%%%%%%%%%%%%%%%%%%%%%%%%%%%%%%%%%%%%%%%%%%%%%%%%%%%%%%%%%%%%%%%%%%%%%%%%

\section{Introduction}

In an integrable model both he spectrum and the eigenstates of the corresponding Hamiltonian, together with the scattering matrix, can be entirely determined by means of any of the different flavours of the Bethe ansatz, such as the coordinate~\cite{Bethe}, 
algebraic~\cite{ABA} or analytic~\cite{Baxter,Reshetikhin} versions of the method. However, this is only half of the journey, as we still have to compute the correlation functions of the model we are examining. On the one hand, this means computing integrals and summing over momentum orderings if we use the coordinate Bethe ansatz. On the other hand, if we use the algebraic Bethe ansatz, we first have to write the local spin operators in terms of the elements of the monodromy matrix. This is the so-called quantum inverse scattering problem~\cite{KMT}. 
If we make use of the commutation relations between the entries of the monodromy matrix, specified by the RTT relations, the correlation functions that involve local operators can be expressed in terms of scalar products of Bethe states 
that usually are off-shell (see for instance~\cite{Kitaninereview} for a review and references therein). Scalar products involving Bethe states normally are long and convoluted, but they simplify in the Heisenberg spin chain when 
at least one of the states is on-shell. 
In this case, they can be expressed in terms of a quotient of determinant~\cite{Korepin,Slavnov}.

The Heisenberg spin chain, originally a toy model of magnetism, is the prototypical example of a quantum integrable system. In recent years, it has played a mayor role in unravelling the integrable structure underlying the AdS/CFT correspondence. 
Thanks to that, today we have a very precise knowledge of the spectrum of local gauge invariant operators ${\cal N}=4$ supersymmetric Yang-Mills. In addition, this lead to some non-trivial tests of AdS/CFT (see for instance~\cite{Beisertreview}). 
In addition, using conformal symmetry, these results are enough to completely fix the two-point function of these operators. However, the problem of determining higher-point correlation function is still not completely solved. 
In recent years, several non-perturbative methods have emerged, such as the bootstrap program~\cite{bootstrap1,bootstrap2} of the hexagon form factor~\cite{hexagon1,hexagon2,hexagon3}. A less refined method to obtain information about higher-point function would be to understanding generic correlation functions 
on the associated spin chain and use them as building blocks. The algebraic Bethe ansatz and the solution to the inverse scattering problem were used in~\cite{RV} to evaluate three-point functions of scalar operators in ${\cal N}=4$ supersymmetric Yang-Mills 
as inner products of Bethe states, constructed out of the elements of the monodromy matrix. As a result, the structure constants of the theory at weak coupling can be written in terms of some elegant determinant 
expressions in~\cite{Escobedo}-\cite{BK}.

If we want to continue this path (for example, to have a source of perturbative data to check the results of the non-perturbative methods), it is desirable a more detailed study of general correlation functions using algebraic Bethe ansatz techniques. 
This article is a pedagogical exercise, where we show how to compute correlation functions involving spin operators located at non-adjacent sites using both the coordinate Bethe ansatz and the algebraic Bethe ansatz. 
The main purpose of this article is to show the subtleties and shortcomings that these kinds of computations present, as some of them are not well known and it is important to be wary of them. One example that is usually not discussed 
is the fact that the states constructed using the coordinate Bethe ansatz and the algebraic Bethe ansatz do differ not only by a normalization factor, but also by a global phase that depends on the momenta of the excitations involved. 
This means that, even if we properly normalise our correlation functions, the ones computed using the coordinate Bethe ansatz still differ from the ones computed using the algebraic one. We will also dedicate space to the apparent singularities 
that arise when naive computing correlation functions using the algebraic Bethe ansatz due to the presence of operators with the same argument.

At the end of this article, we will also include some comments on computations involving the inhomogeneous Heisenberg spin chain. As we have noted above, the Heisenberg model was at the center of the development of integrability 
in the AdS/CFT correspondence. The reason for that was because the conformal dimension of single trace operators on the $SU(2)$ sector of ${\cal N}=4$ supersymmetric Yang-Mills could be described with such model~\cite{MZ}. 
However, the situation is more involved at higher loops, which can only be described using long-range spin chains, meaning interactions beyond nearest-neighbours~\cite{BDS}. However, if we are interested on information up to a certain order 
in the loop expansion and not on an all-loop result, we can use instead a Heisenberg spin chain deformed by the presence of inhomogeneities.~\footnote{Beyond four-loops, it is also necessary to include a dressing phase factor in the S-matrix. Despite that, we will not include it in our computations.}

The remaining part of the article is organised as follows. In section~\ref{Bethe}, we will review the coordinate and the algebraic Bethe ansatz for the isotropic Heisenberg spin chain. 
We include the solution to the inverse scattering problem and a recipe to compute scalar products between an on-shell and an off-shell Bethe state constructed using the Algebraic Bethe Ansatz. In section~\ref{normalizationsection}, 
we discuss the normalisation of states in the coordinate and versions of the ansatz. In section~\ref{correl}, we will explicitly compute correlation functions involving either one or two spin operators, mostly using the algebraic Bethe ansatz. 
The first case is relatively straightforward, but the second one requires some care because the commutation relations of the elements of the monodromy matrix for the homogeneous spin chain seemingly diverge. 
We show that these divergences are only apparent, as they are a $\frac{0}{0}$ indeterminate form that can be treated with proper care. In section~\ref{longcorr}, we will extend the analysis to the long-range BDS spin chain using the fact that it can be mapped to an 
inhomogeneous short-range spin chain. In section~\ref{conclusions}, we will summarise and discuss our results. We close the article with three appendices that collect several additional calculations. 
In appendix~\ref{A} we will solve the recurrence relation~(\ref{recurrenceeq}). In appendix~\ref{B} we will extend our analysis to spin chains 
with $SL(2)$ and $SU(1|1)$ symmetries. In appendix~\ref{C} we will consider the case of correlation functions involving three magnons.

%%%%%%%%%%%%%%%%%%%%%%%%%%%%%%%%%%%%%%%%%%%%%%%%%%%%%%%%%%%%%%%%%%%%%%%%
%%%%%%%%%%%%%%%%%%%%%%%%%%%%%%%%%%%%%%%%%%%%%%%%%%%%%%%%%%%%%%%%%%%%%%%%

\section{The Bethe ansatz}
\label{Bethe}

In this section, we will review some relevant aspects of the coordinate and the algebraic Bethe ansatz for the spin $1/2$ XXX Heisenberg spin chain (see for instance references~\cite{BKI}-\cite{Gomez}). After that, 
we will present a summary of the solution to the inverse scattering problem and compute scalar product, both in the context of the algebraic Bethe ansatz.

\subsection{The coordinate Bethe ansatz}

The Heisenberg model, defined in a one-dimensional closed chain of $L$ fixed particles with spin $1/2$, is given by the Hamiltonian
\begin{equation}
    H=\frac{g^2}{2} \sum_{n=1}^L \left( \mathbb{I}- \sigma^x_n \sigma^x_{n+1} - \sigma^y_n \sigma^y_{n+1} -\sigma^z_n \sigma^z_{n+1}   \right) \ ,
\end{equation}
where $\sigma^i_n$ are the usual Pauli matrices acting on the site (i.e., particle) $n$ of the chain. We are also assuming periodic boundary conditions, which implies the identification $\sigma^i_{L+1}\equiv \sigma^i_1$. 
Another convenient way to write this Hamiltonian is
\be
H = g^2 \sum_{n=1}^L  \big( {\mathbb{I}-\mathbb{P}_{n,n+1}} \big) \ ,
\ee
where $\mathbb{P}_{m,n}$ is the operator that permutes the spin states on positions $m$ and $n$. In the context of ${\cal N}=4$ supersymmetric Yang-Mills (SYM from now on), 
the anomalous dimensions of single trace operators can be mapped to this Hamiltonian, with the following identification of coupling constants~\cite{MZ}
\be
g^2 = \frac{\lambda}{8\pi^ 2} = \frac {g_{\hbox{\tiny{YM}}}^2 N}{8 \pi^2} \ .
\ee

It is immediate to see that the ground state is $L+1$ degenerated. Included among them are the two states with all the spin aligned. Let us pick the state with all the spins up as a starting point for constructing the remaining eigenstates 
(the construction follows the same steps mutatis mutandis if we start from the state with all spins downs). From it, we can define the wavefunctions
\be
\ket{\Psi}=\sum_{1\leq n_1 < n_2 <\cdots < n_N \leq L}{\psi(n_1,\dots ,n_N) \ket{n_1,\dots ,n_N}} \ , \label{CBAstate}
\ee
where $\ket{n_1,\dots ,n_N}$ indicates that the spins in positions $n_i$ have been flipped from their original alignment (i.e., all the spins are up except for those on positions $n_i$, which are spins down). The main assumption 
of the coordinate Bethe ansatz (CBA from now on) is that the above state diagonalises the Heisenberg spin chain provided its wavefunction can be written as the weighted sum of all the possible free wavefunctions we can construct, that is,
\be
\psi(n_1,\dots ,n_N)=\sum_{\sigma\in {\cal P}_N}{A(\sigma,\vec{p}) e^{i(p_{\sigma(1)} n_1+p_{\sigma(2)} n_2+\dots +p_{\sigma(N)} n_N)}} \ , \label{CBAwavefunction}
\ee
where ${\cal P}_N$ is the permutation group of $N$ elements and $A(\sigma,\vec{p})$ is a function that depends on the element of the 
permutation group and the momenta of each magnon, $p_i$. If we substitute the above ansatz into the Schr\"odinger equation $\hat{H} \ket{\Psi}=E \ket{\Psi}$, we obtain the dispersion relation
\be
E = \sum_{j=1}^N{\epsilon(p_j)} \ , \quad \epsilon(p) = 4g^2 \sin ^2 \left( \frac{p}{2} \right) \ ,
\ee
together with the following relation between the weights
\be
A(\sigma_{j,j+1} \sigma , \vec{p}) = S(p_j+1,p_j) A(\sigma , \vec{p}) \ , \quad \hbox{with} \quad S(p,q)=\frac{\frac{1}{2} \cot \left( \frac{p}{2} \right) - \frac{1}{2} \cot \left( \frac{q}{2} \right) + i}{\frac{1}{2} \cot \left( \frac{p}{2} \right) - \frac{1}{2} 
\cot \left( \frac{q}{2} \right) - i} \ ,
\label{Spq}
\ee
where $\sigma_{j,j+1}$ is the permutation that swaps positions $j$ and $j+1$ while leaving the remaining ones fixed. The function $S(p,q)$ is called S-matrix of the Heisenberg model. Thus, the wavefunction for the two-magnon state takes the form
\be
\psi(n_1,n_2) = e^{i(p_1 n_1 + p_2 n_2)} + S(p_2,p_1) e^{i(p_2 n_1 + p_1 n_2)} \ , \label{twomagnonwavefunction}
\ee
up to the global normalization factor $A(\mathbb{I} , \vec{p})$, which we will set to 1.

However, not every arbitrary set of momenta will lead to a wavefunction that satisfy the correct periodic boundary conditions. Periodicity imposes $N$ quantization conditions, one for each of the momenta, known as Bethe equations,
\be
e^{ip_j L} \prod_{k\neq j}^N{S(p_k,p_j)}=1 \ .
\label{BAECBA}
\ee
The physical meaning of this equation is that, if we carry one magnon with momentum $p_j$ around the chain, the free propagation phase $p_j L$ 
plus the phase change due to the scattering with each of the other $N-1$ magnons must give a trivial phase.

If we plan to apply this machinery to compute the one-loop correction to the dimension of single-trace operators in ${\cal N}=4$ SYM theory, we need to impose an additional condition on the momenta. 
Periodicity of the spin chain implies that it is invariant under the simultaneous shift of all positions by the length of the chain, $n_i \rightarrow n_i + L$, but cyclicity of the trace demands the stronger condition 
of invariance under the simultaneous shift of all positions by one site, $n_i \rightarrow n_i + 1$. This means that our states have to satisfy the zero total momentum condition
\be 
\prod_{i=1}^N{e^{ip_j}}=1 \ .
\label{trace}
\ee

\subsection{The algebraic Bethe ansatz}

Let us now move to briefly review the algebraic Bethe ansatz (ABA from now on) for the inhomogeneous spin~$1/2$ Heisenberg spin chain. We will 
mostly follow notation and conventions in reference~\cite{KMT}. The core of the ABA is the quantum R-matrix, which is an operator that satisfies 
the Yang-Baxter equations. In the Heisenberg chain it is given by
\be
R(\lambda , \mu) = \left( \begin{matrix}
1 & 0 & 0 & 0 \\
0 & b(\l,\m) & c(\l,\m) & 0 \\
0 & c(\l,\m) & b(\l,\m) & 0 \\
0 & 0 & 0 & 1 
\end{matrix} \right) \ ,
\ee
where $b(\l,\m)$ and $c(\l,\m)$ are functions of the rapidities $\l$ and $\mu$, 
\be
b(\l,\m) = \frac {\l - \m}{\l - \m + \eta} \ , \quad c(\l,\m) =  \frac {\eta}{\l - \m + \eta} \ .  
\ee
In these functions $\eta$ is the so-called crossing parameter, which we will take to be $i$. The monodromy matrix of the spin chain is constructed as an ordered product of R-matrices, 
\be
T_{0}(\l) =\prod_{n=L}^1{R_{0,n}} = R_{0L}(\l,\xi_{L}) \ldots R_{01}(\l,\xi_{1}) \ ,
\ee
where the labels $1$ to $L$ represent the physical sites of the spin chain and the label $0$ represents an additional auxiliary space. 
The variables $\xi_n$ are called the inhomogeneities of the spin chain. For the case of the Heisenberg spin chain, which is the case that we are going to study during 
most of this article, we will consider the homogeneous case $\xi_n = \xi = \frac{\eta}{2}$ for any value of $n$. The monodromy matrix, represented as a $2 \times 2$ matrix in auxiliary space, takes the form
\begin{align}
	T_0(\lambda)&= \left( \begin{matrix}
	A(\lambda) & B(\lambda) \\
	C(\lambda) & D(\lambda) \end{matrix} \right) \ .
\label{transfer}
\end{align}

The trace of the monodromy matrix, known as the transfer matrix, has two very important properties: it commutes with itself for different values of $\lambda$, and it is the generating function of the conserved charges 
of the Heinsenberg spin chain. Therefore, by diagonalising the transfer matrix, we are simultaneously diagonalising all the conserved charges. In the ABA, the diagonalisation process is not done brute-force, 
but indirectly by using the commutation relations of the entries of the monodromy matrix. First, we assume the existence of a pseudo-vacuum $|0\rangle$. In the Heisenberg chain this is just the completely ferromagnetic 
state, with all spins up, and can be represented by the tensor product 
\be
\quad \ket{0}=\bigotimes_{n=1}^L \ket{0}_n \ , \quad \hbox{where} \quad \ket{0}_n = \left( \begin{matrix} 1  \\ 0 \end{matrix} \right)_n \ .
\ee
The elements of the monodromy matrix act on the pseudo-vacuum as
\be
A(\l) |0 \rangle = a(\l) |0 \rangle \ , \quad D(\l) |0 \rangle = d(\l) |0 \rangle \ , \quad C(\l) |0 \rangle = 0 \ .
\ee
With the above choice of R-matrix, the eigenvalues $a(\l)$ and $d(\l)$ take the values\footnote{For generic values of the inhomogeneities, the  function $d$ is given by
\[
d(\lambda)=\prod_{n=1}^L{\frac{(\lambda-\xi_n)}{(\lambda-\xi_n+\eta)}} \ ,
\]
instead. Notice that it always vanishes when $\lambda$ is equal to the value of one of the inhomogeneities, $d(\xi_n)=0$ for any $n$.}
\be
a(\lambda)=1\ , \quad 
d(\lambda)=\frac{(\lambda-\xi)^L}{(\lambda+\xi)^L} \ .
\ee
The commutation relations between elements of the monodromy matrix are fixed by the so called RTT relations
\begin{equation}
 R_{12} (\lambda , \mu ) T_1 (\lambda) T_2 (\mu) =T_2 (\mu) T_1 (\lambda) R_{12} (\lambda, \mu ) \ .
\end{equation}
Of the commutation relations encoded on this equation, we are mostly interested on~\cite{Korepin,Faddeev},
\begin{align}
	(A+D) (\mu) B(\lambda) &= B(\lambda) (A+D) (\mu) -\frac{\eta}{\lambda-\mu} \left[ B(\lambda) (D-A)(\mu) -B(\mu) (D-A) (\lambda) \right] \ , \label{commBAD} \\
	C(\lambda) (A+D) (\mu) &= (A+D) (\mu) C(\lambda) -\frac{\eta}{\lambda-\mu} \left[ (D-A)(\mu) C(\lambda) -(D-A) (\lambda) C(\mu) \right] \ , \label{commCAD}  \\
	\left[C(\lambda), B(\mu) \right] &=-\frac{\eta}{\lambda-\mu} \left( A(\lambda) D(\mu)-A(\mu) D(\lambda) \right) \ . \label{commCB}
\end{align}
Note that these commutation relations lead to an indeterminate form of the type $\frac{0}{0}$ when $\l$ and $\m$ are equal. This will later lead to some issues when computing correlation functions.

A natural procedure to construct the eigenstates of the transfer matrix is to apply the operator $B(\l)$ on the pseudo-vacuum. 
Using the previous commutation relations, one can find the action of the operator $(A+D)(\mu)$ on $B(\l) | 0 \rangle$ and impose that these 
states must be eigenstates of that operator, 
\be
	(A+D)(\mu ) \prod_{i=1}^N{B(\l _i )} \ket{0}=\tau (\mu,\{ \l \})\prod_{i=1}^N{B(\l _i}) \ket{0} \:\:
	\Rightarrow \:\: \frac{a(\l _i)}{d(\l _i)} \prod_{j\neq i}{\frac{\l _i-\l _j-i}{\l _i -\l _j +i}}=1 \ ,
\label{BAEABA}
\ee
which are the Bethe equations for the ABA. If we want to recover the Bethe equations derived in the previous subsection from the CBA, we need to write 
the momenta of the magnons as a function of the rapidity, \footnote{We follow the definition of the momentum in~\cite{KMT}. In the literature, it is also common the opposite definition of momentum, 
as both the CBA ansatz and the Bethe equations are invariant under the transformation $p\rightarrow -p$ and $x\rightarrow -x$.}
\be 
\frac{\l _j -\xi}{\l _j+\xi}=e^{ip_j} \longleftrightarrow \lambda(p) =-\frac{1}{2} \cot \left( \frac{p}{2} \right) \label{momentum} \ .
\ee
If we perform this substitution, the S-matrix becomes
\be
S_{ij} = \frac {\l_j - \l_i + i}{\l_j - \l_i - i} \ ,
\ee
and equation (\ref{BAEABA}) agrees with the Bethe equations (\ref{BAECBA}) in the CBA. 

As we commented above, the Heisenberg Hamiltonian is connected to the one-loop anomalous dimension of single trace operators in $\mathcal{N}=4$ SYM. We can extend this spin chain picture to higher loops, but it requires the Hamiltonian 
to be of long range. In fact, the range of the interaction is, at most, equal to the loop order we want to describe. Luckily, we can circumvent the need for the long-range interaction thanks to the inhomogeneities 
(at least, up to an order equal to the length of the spin chain). This is due to the fact that the long-range spin chain that describes this situation can be modelled by a similar set of Bethe equations
\be 
e^{ip_i L} \prod_{j\neq i}^N {\frac{\l(p_i)-\l(p_j)+i}{\l(p_i)-\l(p_j)-i}} =1 \ ,
\ee
where the rapidities are deformed to~\cite{BDS}
\be
\lambda(p) =-\frac{1}{2} \cot \left( \frac{p}{2} \right) \sqrt{1 + 8g^2 \sin^2 \left( \frac {p}{2} \right)} \ .
\label{longrangerapidity}
\ee
Inverting this relation,  we can present these equations in a more convenient way 
\be
\frac {x(\l_i+i/2)^L}{x(\l_i-i/2)^L} = \prod_{j\neq i}^N \frac {\l_i - \l_j + i}{\l_i - \l_j - i} \ , 
\ee
where $x(\l)$ is given by
\be
x(\l) = \frac {1}{2} \l + \frac {1}{2} \sqrt{\l^2 - 2g^2} \ .
\ee
Thus, if we are interested in the correction up to $L$-loops, we can model the homogeneous long-range spin chain by an inhomogeneous nearest-neighbours spin chain
\be
\frac {P_L(\l_i + i/2)}{P_L(\l_i - i/2)} = \prod_{j\neq i}^N \frac {\l_i - \l_j + i}{\l_i - \l_j - i} \ , 
\ee
where the polynomial $P_L(\l)$ is given by
\be
P_L(\l) = \prod_{n=1}^L (\l - \xi_n) \ , \quad \hbox{with} \quad \xi_n = \frac{i}{2} +\sqrt{2} g\cos\frac{(2n-1)\pi}{2L} \ .
\ee

%%%%%%%%%%%%%%%%%%%%%%%%%%%%%%%%%%%%%%%%%%%%%%%%%%%%%%%%%%%%%%%%%%%%%%%%
%%%%%%%%%%%%%%%%%%%%%%%%%%%%%%%%%%%%%%%%%%%%%%%%%%%%%%%%%%%%%%%%%%%%%%%%

\subsection{Inverse scattering and inner products of Bethe states}

Both the CBA and the ABA constitute only half of the work we have to do to solve a theory. They provide us with the set of states and their energy, but they do not give us a recipe to compute the expectation value of operators. 
For the CBA, which is more akin to first quantisation, we can easily compute the action of operators on our wavefunctions and, with a little more effort, compute the scalar product of states. In contrast, the ABA is more akin to second quantisation, 
so we do not have direct access to the wavefunctions as it is not immediately evident how the operators $A$, $B$, $C$ and $D$ relate to the spin operators actin on a given site. In this section, we discuss these two points.

\subsubsection*{The inverse scattering problem}

First, we will discuss how to relate the entries of the monodromy matrix and the local spin operators appearing in the CBA. These relations were first found in~\cite{KMT} using the method of factorizing F-matrices 
for the case of the general XXZ inhomogeneous spin chain and later in~\cite{GK} for the XYZ homogeneous spin chain using the properties of the R-matrix and the monodromy matrix. The general computation is lengthy and cumbersome, 
so we will only prove here the relation in the homogeneous case, where it simplifies, and just present the final result for the inhomogeneous case.

The key point to match the entries of the monodromy matrix and the local spin operators is to realise that the monodromy matrix at $\lambda=\xi$ can be written as
\begin{equation}
    T_0(\xi)=\mathbb{P}_{01} \mathbb{P}_{1L}  \mathbb{P}_{1,L-1} \dots \mathbb{P}_{13} \mathbb{P}_{12}= \mathbb{P}_{01}  U \ ,
\end{equation}
that is, $\mathbb{P}_{01} $ times the operator that shifts all positions by one to the right. This means that $A(\xi)+D(\xi)$, $B(\xi)$, $C(\xi)$ and $A(\xi)-D(\xi)$ are in one to one correspondence 
with $\mathbb{I}$, $\sigma^-_1$, $\sigma^+_1$ and $\sigma^z_1$ respectively. We only have to compensate the additional $U$ factor by multiplying by $U^{L-1}$ on the right. 
Operators acting on the remaining sites of the chain can be retrieved by the adjoint action of $U$ (or, equivalently, $A(\xi)-D(\xi)$) an appropriate number of times.

The solution for the inhomogeneous spin chain is, surprisingly, not much different
\begin{align}
	\sigma^+_k &=\prod_{i=1}^{k-1}{ (A+D) (\xi_i)} \, C(\xi_k) \, \prod_{i=k+1}^L{ (A+D) (\xi_i)} \ , \label{sigmaplus} \\
	\sigma^-_k &=\prod_{i=1}^{k-1}{ (A+D) (\xi_i)} \, B(\xi_k) \, \prod_{i=k+1}^L{ (A+D) (\xi_i)} \ , \label{sigmaminus} \\
	\sigma^z_k &=\prod_{i=1}^{k-1}{ (A+D) (\xi_i)} \, (A-D)(\xi_k) \, \prod_{i=k+1}^L{ (A+D) (\xi_i)} \ . \label{sigmaz}
\end{align}
Thus, we can compute expectation values of local operators without using nothing but the Yang-Baxter algebra.

\subsubsection*{Scalar products}
\label{howtocomputescalarproducts}

The main issue with the above expressions for local operators are the additional $C$ and $B$ operators they introduce, as the new states we will have to work with no longer satisfy the Bethe equations. 
This is a difficult topic and there is a large amount of literature devoted to this kind of computation (see, for example,~\cite{BKI} and references therein). The expressions we get are usually very involved and difficult to handle, 
but they simplify heavily when the set of rapiditities that describes one of the states fulfils the Bethe equations. Given the scalar product
\be
	S_N (\{ \mu_j \} , \{\lambda_k \})=S_N (\{\lambda_k \} , \{ \mu_j \})=\langle 0 | \prod_{j=1}^N C(\mu _j) \prod_{k=1}^N B(\l _k) | 0 \rangle \ ,
\ee
where the set of rapidities $\{ \l _k\}$ is a solution to the Bethe equations and $\{ \mu _j \}$ is an arbitrary set of parameters, it can be represented as a ratio of two determinants, 
\be 
S_N (\{ \mu_j \} , \{\lambda_k \})=\frac{\det T}{\det V} \ , 
\label{scalarproduct}
\ee
where $T$ and $V$ are $N\times N$ matrices given by
\begin{align}
T_{ab} &=\dpartial{\tau (\mu_b,\{ \l \})}{\lambda_a} \ , & &\tau (\mu,\{ \l \}) 
=a(\mu) \prod_{k=1}^N{\frac{\l _k-\mu+i}{\l_k -\mu}}+d(\mu) \prod_{k=1}^N{\frac{\l _k-\mu-i}{\l_k -\mu}} \ , \notag\\
V_{ab} &=\frac{1}{\mu_b-\l _a} \ , & &\det V =\frac{\prod_{a<b}{(\l _a -\l _b)} \prod_{j<k}{(\mu_k-\mu_j)}}
{\prod_{k=1}^N{\prod_{a=1}^N{(\mu_k -\lambda_a)}}} \ . 
\label{howtoscalarproduct}
\end{align}
Similar expressions hold for the case where the set $\{ \mu _j \}$ is a solution to the Bethe equations and $\{ \l _k\}$ is an arbitrary set of parameters.

If we take the limit $\mu_a \rightarrow \lambda_a$ in these expressions, we recover the Gaudin formula for the square of the norm of a Bethe state~\cite{GaudinSP,Korepin},
\begin{gather}
S_N (\{\lambda_k \}, \{\lambda_k \})= 
i^N \prod_{j \neq  k}{\frac{\l _j -\l _k +i}{\l _j -\l _k}} \det \Phi '(\{\lambda_k \}) \ , \nonumber \\
\Phi '_{ab} (\{\lambda_k \}) =-\dpartial{}{\l _b} \ln \left( \frac{a (\l _a)}{d(\l _a)} \prod_{b\neq a}{\frac{\l _a-\l _b+i}{\l _a -\l _b -i}} \right) \ .
\label{gaudin}
\end{gather}
We should stress that this way of calculating scalar products is valid for the case of a finite spin chain. 
The generalization of these expressions to the thermodynamical limit of very long chains can be found, for example, 
in~\cite{Maillet}.

%%%%%%%%%%%%%%%%%%%%%%%%%%%%%%%%%%%%%%%%%%%%%%%%%%%%%%%%%%%%%%%%%%%%%%%%
%%%%%%%%%%%%%%%%%%%%%%%%%%%%%%%%%%%%%%%%%%%%%%%%%%%%%%%%%%%%%%%%%%%%%%%%

\section{Normalization of states}
\label{normalizationsection}

States constructed using the CBA are normalised differently from the ones constructed using the ABA \cite{Sklyanin:1980ij}. Thus, before computing generic correlation functions, we have to discuss how these states differ.

The simplest correlation function where we can see a discrepancy between the two normalisations is  $\granesperado{\lambda}{\sigma^+_k \sigma^-_l}{\lambda}$. 
In the CBA, this correlation function is simply given by $e^{ip(l-k)}$. In order to perform the computation in the ABA, we first have to write the spin operators in terms of elements of the monodromy matrix, 
\ba
\granesperado{\lambda}{\sigma^+_k \sigma^-_l}{\lambda} & \!\! = \!\! & \granesperado{0}{C(\lambda) \, 
(A+D)^{k-1} \, B(\xi) \, (A+D)^{L-k+l-1} \, C(\xi) \, (A+D)^{L-l} \, B(\lambda)}{0} \nonumber \\
& = \!\! & e^{-ip(L-l+k-1)} \granesperado{0}{C(\lambda) \,  B(\xi) \, (A+D)^{L-k+l-1} \, C(\xi) \, B(\lambda)}{0} \ .
\ea
From the commutation relations~(\ref{commCB}) we find that
\be
\bra{0} C(\lambda) \, B(\xi)=i\frac{d(\lambda)}{\lambda-\xi} \bra{0} \ , 
\ee
with an identical result for $C(\xi) \, B(\lambda) \ket{0}$. Recalling that the Bethe equation for the single-magnon state reads $d(\lambda)=1$ we conclude that 
\be
\granesperado{\lambda}{\sigma^+_k \sigma^-_l}{\lambda} = \frac{i^2 e^{ip(l-k+1)}}{(\l-\xi)^2} \ .
\ee
We can try to solve the disagreement with the CBA by dividing this result by the norm of the state. 
This can be easily computed using the Gaudin formula~(\ref{gaudin}), 
\be
\langle \lambda | \lambda \rangle =i\dpartial{d}{\lambda}=\frac{i^2 L}{\lambda^2-\xi^2} \ . 
\ee
Therefore
\begin{align}
	& \frac {\granesperado{\lambda}{\sigma^+_k \sigma^-_l}{\lambda}}{\langle \lambda | \lambda \rangle}
	= \frac {e^{ip(l-k)}}{L} \left( \frac{\lambda+\xi}{\lambda-\xi} e^{ip} \right) = \frac {e^{ip(l-k)}}{L} \ , 
\end{align}
which is the result in the CBA provided we normalise it.

However, the disagreement between both wavefunctions cannot be cured only by dividing by the norm of the states involved, as they also disagree by a global phase. 
The simplest case where we can see this discrepancy is $\granesperado{0}{\sigma^+_k}{\lambda}$, i.e., the one magnon wavefunction
\be
\frac{\granesperado{0}{\sigma^+_k}{\lambda}}{\sqrt{\langle \lambda | \lambda \rangle}}=
\frac{e^{ipk}}{\sqrt{L}} \ \sqrt{\frac{\lambda+\xi}{\lambda-\xi}}=\frac{e^{ip(k-\frac{1}{2})}}{\sqrt{L}} \ .
\ee
Thus, the states computed using the ABA also differ from the ones computes using the CBA by a global phase that depends on the rapidity.

In order to fix the normalization of states in the ABA with respect to the normalization of states in the CBA, 
we should go back to the definition of the transfer matrix, equation (\ref{transfer}),
and apply it to the ground state,
\begin{align}
	& R_{0,n} \ket{0}_n =\left( \begin{matrix}
	1 & \frac{i}{\lambda-\xi+i} \sigma^-_n \\
	0 & \frac{\lambda-\xi+i}{\lambda-\xi} \end{matrix} \right) \ket{0}_n \ .
\end{align}
If we focus on the operator $B(\lambda)$, we can write
\ba
B(\lambda) & \!\! = \!\!  & \frac{i}{\lambda+\xi} \sum_{n=1}^L{ \sigma^-_n \left( \frac{\lambda-\xi}{\lambda+\xi} \right)^{n-1} } + \dots
= \frac{i}{\lambda-\xi} \sum_{n=1}^L{ \sigma^-_n e^{ipn}} + \dots \ ,
\ea
where the dots represent terms that vanish when $B(\lambda)$ is applied to the ground state. Therefore, states with a single magnon in the ABA, $\ket{\lambda}^{\text{a}}$, relate to states in the CBA through
\be
B(\lambda)\ket{0}=\ket{\lambda}^{\text{a}}=\frac{i}{\lambda-\xi} \ket{\lambda}^{\text{c}} \ . 
\ee
We can repeat it with the dual ground state
\be
_n \! \bra{0} R_{0,n} =  \, _n \! \bra{0} \left( \begin{matrix}
\frac{\lambda-\xi+i}{\lambda-\xi} & 0 \\
\frac{i}{\lambda-\xi+i} S^+_n & 1 \end{matrix} \right)	\ , 
\ee
from where we conclude that
\be
^{\text{a}} \bra{\lambda}=i\frac{d(\lambda)}{\lambda+\xi} \  ^{\text{c}} \! \bra{\lambda} \ , 
\ee
and therefore
\ba
C(\lambda) & \!\! = \!\! & \frac{i\, d(\lambda)}{\lambda+\xi} \sum_{n=1}^L{ S^+_n \left( \frac{\lambda-\xi}{\lambda+\xi} \right)^{-n} + \dots} 
= \frac{i\, d(\lambda)}{\lambda+\xi} \sum_{n=1}^L{ S^+_n e^{-ipn}} + \dots \ .
\ea
An identical discussion holds in the case of states with more than one magnon, giving us the general relations
\begin{gather}
	\ket{\l _1 , \l _2 , \dots , \l _N}^{\text{a}}= \prod_{j=1}^N{ \frac{i}{(\l _j-\xi )}}
	\prod_{i<j}{\frac{\l _j-\l _i+i}{\l _j-\l _i}} \, \ket{\l _1 , \l _2 , \dots , \l _N}^{\text{c}} \ , \\	
	\bra{\l _1 , \l _2 , \dots , \l _N}^{\text{a}}= \prod_{j=1}^N{ i \frac{d(\l_j )}{(\l _j+\xi )}}
	\prod_{i<j}{\frac{\l _j-\l _i-i}{\l _j-\l _i}} \, \bra{\l _1 , \l _2 , \dots , \l _N}^{\text{c}}\ .
\end{gather}
The first product can be removed by an appropriate normalization of the states, and thus there will only remain a shift 
in the position of the coordinates by $-\frac{1}{2}$. The second factor is related to the fact that CBA states 
are not symmetric if we interchange two magnons. In fact, they pick up a phase which is equal to the S-matrix. 
On the other hand, ABA states are symmetric under the exchange of two magnons. A nice intermediate choice is to fix this phase in such a way that the correlation functions 
have the structure $\sqrt{\prod_{\mu_i<\mu_j}{S_{ij}}}\cdot \{\text{term symmetric in the rapidities}\}$.
Despite looking like a very ad hoc solution, we will justify this choice below.

An alternative normalisation of the ABA can be obtained if, instead of using B-states to define the excitations in the ABA, we use Z-states, defined as
\begin{equation}
	Z(\lambda)=B(\lambda) A^{-1} (\lambda) \ .
\end{equation}
In fact it is natural to use these states because they generate a Zamolodchikov-Faddeev algebra~\cite{Faddeev}, 
\be
Z(\l) Z(\m)=Z(\m) Z(\l) S_{\m \l}=Z(\m) Z(\l) \frac{\m -\l -i}{\m -\l +i} \ . 
\ee
Using these operators as creation operators, states constructed using the ABA behave as states constructed using the CBA under the exchange of two magnons.

We should point that this relation is only true in the large length limit $L\rightarrow \infty$. To see that, let us consider again the commutation relation between $A$ and $B$
\begin{align}
B(\m) A(\l) &= \left( 1+ \frac{i}{\l-\m} \right) A(\l) B (\m) - \frac{i}{\l-\m} A(\m) B (\l) \label{commAB} \ .
\end{align}
At large $L$ we can effectively disregard the second term in the right-hand-side of the commutation relation. Consequently, the relationship between the $Z$-states and the $B$-states acting on the ground state in this limit is given by
\begin{equation}
\mathcal{R} \left[ \prod_i{Z(\mu_i) \ket{0}} \right] = \prod_{i<j}{\frac{\mu_j-\mu_i}{\mu_j-\mu_i+i}} 
\prod_i{B(\mu_i) \ket{0}} \ ,
\end{equation}
where $\mathcal{R}$ indicates that we are ordering the $Z$ operator such that $\mu_j$ appears to the right of $\mu_i$ if $j>i$. Hence, using the Zamolodchikov-Faddeev states instead of the usual magnon states introduces a phase shift. In fact, this is precisely the phase we wanted to introduce ad hoc before.

Similarly, we can define the operators $F(\lambda)=d(\lambda) \, D^{-1} (\lambda) \, C(\lambda)$ as annihilation operators. The relationship between correlation functions computed in the $F$ and $Z$ and correlation functions computed in the $C$ and $B$ basis is given by
\begin{equation}
	\granesperado{0}{F(\m) F(\l) \dots Z(\l) Z(\m)}{0}= \, 
	\frac{(\m -\l)^2 \dots}{(\m -\l -i)(\m -\l +i) \dots} \granesperado{0}{C(\m) C(\l) \dots B(\l) B(\m)}{0} \ , \notag
\end{equation}
which is symmetric under exchange of $\l$ and $\m$, as we wanted. We should stress again that this relation is valid only in the $L\rightarrow \infty$ limit.

%%%%%%%%%%%%%%%%%%%%%%%%%%%%%%%%%%%%%%%%%%%%%%%%%%%%%%%%%%%%%%%%%%%%%%%%%
%%%%%%%%%%%%%%%%%%%%%%%%%%%%%%%%%%%%%%%%%%%%%%%%%%%%%%%%%%%%%%%%%%%%%%%%%

\section{Correlation functions} \label{correl}

We will now apply the tools introduced in the previous section to evaluate correlation functions of spin operators, both using the CBA and the ABA, in order to compare the corresponding difficulties arising in each method.

\subsection{Correlation functions involving two operators in the CBA}

Let us start with a discussion on the computation correlation functions in the context of the CBA. In general, the correlation functions that can be analysed are of the form
\begin{equation}
    \langle \Psi ' | \sigma^-_{k_1} \dots \sigma^-_{k_m} \sigma^+_{l_1} \dots \sigma^+_{l_n} | \Psi \rangle \ ,
\end{equation}
but in order to face the difficulties arising in the calculation it will suffice to consider just the case of $m=n=1$. In this case, we can use the invariance under simultaneous translations 
of all the positions to set $k_1=L$, which simplifies the form of the wavefunctions involved. Substituting the states (\ref{CBAstate}), the correlation function takes the form
\begin{equation}
    \sum_{j=1}^{N-1}\sum_{\{n_i\}}{[\psi ' (n_1,\dots ,n_{N-1}, L)]^* \psi(n_1,\dots , n_{j-1} , l_1, n_{j}, \dots ,n_{N-1})} \ ,
\end{equation}
where the second sum is performed over the positions $n_i$, with the restriction $1\leq n_1 < n_2 <\cdots < n_{j-1} < l_1<n_{j}< \dots < n_{N-1} \leq L$. To obtain this expression, 
we have used that $\sigma^-$ (respectively $\sigma^+$) annihilates one of the excitations of $| \Psi \rangle$ (respectively $\langle \Psi ' |$) and fixes one of the positions. 
Once that is done, orthogonality of the states $\ket{n_1,\dots ,n_N}$ forces the remaining positions of the excitations of both wavefunctions to be the same. 
If we substitute now the wavefunctions (\ref{CBAwavefunction}), the previous equation becomes
\begin{multline}
    \sum_{j=1}^{N-1}\sum_{\sigma\in \mathcal{P}_N } \sum_{\sigma'\in \mathcal{P}_{N-1} }\sum_{\{n_i\}} {A^*(\sigma',\vec{p}') A(\sigma,\vec{p})  e^{i [p_{\sigma(j)}l_1 -  p'_{\sigma'(N)} L] }} \ \cdot \\
    e^{i[(p_{\sigma(1)}-p'_{\sigma'(1)}) n_1+\dots +(p_{\sigma(j-1)}-p'_{\sigma'(j-1)}) n_{j-1} + (p_{\sigma(j+1)}-p'_{\sigma'(j)}) n_j +\dots +(p_{\sigma(N)}-p'_{\sigma'(N-1)}) n_{N-1}} \ . \label{esperadoCBA}
\end{multline}
In the large $L$ regime, we can substitute the sum over positions by a set of nested integrals,
\begin{equation}
    \sum_{\{n_i\}} e^{i p_i n_i} \approx \int_0^L dx_{N-1} \int_0^{x_{N-1}} dx_{N-2} \dots \int_0^{x_2} dx_1 e^{i p_i x_i} \ .
\end{equation}
This simplifies the computations, as each integral is of the form ``polynomial times exponential''. However, a closed form for this integral does not exist, as its form changes radically depending on how many 
and which $p_i$ vanish (as they give rise to the polynomial part of the integrand). Furthermore, once the integrals have been evaluated, we still have to deal with the combinatorial problem 
that the sums over the permutations $\sigma$ and $\sigma '$ create. This is far from trivial, even in the simple case when we were computing just an inner product (i.e., $m=n=0$), as can be seen 
for instance in~\cite{GaudinSP}. In the case of larger values of $m$ and $n$, the complexity of the integrals that have to be addressed remains the same, but the complexity of the combinatorial problem 
increases significantly (specially if $n\neq m$), because the sum to be performed contains several ordering restrictions.

\subsection{Correlation functions involving one operator in the ABA}
\label{lambdasigmamumusection}

We will now move to the computation of correlation functions using the ABA. We will begin by computing the case of the three-magnon correlation function $\granesperado{\lambda}{\sigma^+_k }{\mu_1 \mu_2}$ 
(the extension to correlators with $2n-1$ magnons is immediate and we will not present it here). 
Using relation (\ref{sigmaplus}) we can bring the problem to a calculation in the ABA,
\ba
\granesperado{\lambda}{\sigma^+_k }{\mu_1 \mu_2}^{\text{a}} \!\! & = & \!\! 
\granesperado{0}{C(\lambda) \, (A+D)^ {k-1} (\xi) \, C(\xi) \, (A+D)^ {L-k} (\xi) \, B(\mu_1) \, B(\mu_2)}{0} \nonumber \\
& = & \!\! e^{-i [(p_1+p_2)\cdot (L-k)+p_{\lambda}(k-1)]} \granesperado{0}{C(\lambda) \,  C(\xi) \, B(\mu_1) \, B(\mu_2)}{0} \ .
\ea
Note that although $\l$ satisfies the Bethe equations for a single-magnon state, the pair $\{ \l, \xi\}$ does not define a Bethe state. 
Therefore, we need to calculate the scalar product of an arbitrary vector with a Bethe state. 
This can be done following the recipe we stated in section~\ref{howtocomputescalarproducts}. % The first step is to write (recall that $\xi=i/2$ for the Heisenberg chain)
%\begin{align}
%	\tau (\xi,\{\mu_1 , \mu_2\} ) &=\frac{\mu_1-\xi+i}{\mu_1-\xi} \, \frac{\mu_2-\xi+i}{\mu_2-\xi} 
%	= \frac{\mu_1+\xi}{\mu_1-\xi} \, \frac{\mu_2+\xi}{\mu_2-\xi} \ , \nonumber \\
%	\tau (\lambda,\{\mu_1 , \mu_2\} ) &=\frac{\mu_1-\lambda+2\xi}{\mu_1-\lambda} \, \frac{\mu_2-\lambda+2\xi}{\mu_2-\lambda}+d(\lambda) \, 
	%\frac{\mu_1-\lambda-2\xi}{\mu_1-\lambda} \, \frac{\mu_2-\lambda-2\xi}{\mu_2-\lambda} \ ,
%\end{align}
%so that the $T$ and $V$ matrices are given by
For these states, the matrices $T$ and $V$ take the form (recall that $\xi=i/2$ for the Heisenberg chain)
\begin{align}
	T_{11} &	= \frac{-2\xi}{(\mu_1-\xi)^2} \frac{\mu_2+\xi}{\mu_2-\xi} \ , \quad 
	T_{21} = \frac{\mu_1+\xi}{\mu_1-\xi} \frac{-2\xi}{(\mu_2-\xi)^2} \ , \nonumber \\
	T_{12} &=\frac{-2\xi}{(\mu_1-\lambda)^2} \, \frac{\mu_2-\lambda+2\xi}{\mu_2-\lambda}+ \frac{2\xi}{(\mu_1-\lambda)^2} \, 
	\frac{\mu_2-\lambda-2\xi}{\mu_2-\lambda} \ , \nonumber \\
	T_{22} &= \frac{\mu_1-\lambda+2\xi}{\mu_1-\lambda} \, \frac{-2\xi}{(\mu_2-\lambda)^2}+\frac{\mu_1-\lambda-2\xi}{\mu_1-\lambda} \, 
	\frac{2\xi}{(\mu_2-\lambda)^2} \ , \nonumber \\
	\frac{1}{V} &=\frac{(\mu_1 -\xi) (\mu_1 -\lambda)(\mu_2 -\xi)(\mu_2 -\lambda)}{(\lambda-\xi) (\mu_1 -\mu_2)} \ ,
\end{align}
where we have used that the Bethe equations imply $d(\lambda)=1$ for the case of a single-magnon. 
After some immediate algebra, the correlation function becomes
\begin{equation}
	\granesperado{\lambda}{\sigma^+_k }{\mu_1 \mu_2}^{\text{a}}=\frac{16\xi^3 \, e^{i(p_1+p_2-p_\lambda )k}}{(\lambda+\xi) (\mu_1 -\mu_2)} \, 
	\left[ \frac{\mu_2+\xi}{(\mu_1-\xi)(\mu_2-\lambda)} -\frac{\mu_1+\xi}{(\mu_2-\xi)(\mu_1-\lambda)} \right] \ .
\end{equation}
Now, if we want to read this result in the normalization of the CBA, we need to recall the discussion in section~\ref{normalizationsection}. In the case at hand
\be
\granesperado{\lambda}{\sigma^+_k}{\mu_1 \mu_2}^{\text{a}}=\frac{i\, d(\lambda)}{\lambda+\xi} \, 
\frac{\mu_1-\mu_2-i}{\mu_2-\mu_1} \, \frac{1}{(\mu_1-\xi)(\mu_2-\xi)} \granesperado{\lambda}{\sigma^+_k}{\mu_1 \mu_2}^{\text{c}} \ .
\ee
Therefore
\begin{equation}
	\granesperado{\lambda}{\sigma^+_k}{\mu_1 \mu_2}^{\text{c}}=e^{i(p_1+p_2-p_\lambda )k}\frac{2}{\mu_1-\mu_2- i} \, 
	\left[ \frac{\mu_2^2-\xi^2}{(\mu_2-\lambda)} -\frac{\mu_1^2-\xi^2}{(\mu_1-\lambda)} \right] \ .
\end{equation}
Let us divide by the norm of the states in both cases,
which can be easily calculated using the Gaudin formula~(\ref{gaudin}). 
In the ABA,
\be
\braket{\mu_1 ,\mu_2}{\mu_1 , \mu_2}^{\text{a}} = \frac{16\xi ^4 L^2 \, \big[ (\mu_2 -\mu_1)^2 -4\xi^2 \big]}{(\mu_2-\mu_1)^2 
    \left( \mu_1^2 -\xi^2 \right) \left( \mu_2^2 -\xi^2 \right)} \left( 1 -\frac{2}{L} \cdot 
    \frac{\left( \mu_1^2 +\mu_2^2 -2\xi^2 \right)}{\left[ (\mu_2 -\mu_1)^2 -4\xi^2 \right]} \right) \ . 
\label{algebraicnorm}
\ee    
Recalling again section~\ref{normalizationsection}, states in the CBA and the ABA  are related through 
\be
\braket{\mu_1 ,\mu_2}{\mu_1 , \mu_2}^{\text{a}} = \left( \frac{\mu_2-\mu_1+i}{\mu_1-\mu_2} \right) \left( \frac{\mu_2-\mu_1-i}{\mu_1-\mu_2} \right) 
\frac{\braket{\mu_1 ,\mu_2}{\mu_1 , \mu_2}^{\text{c}} }{{\left( \mu_1^2 -\xi^2 \right) \left( \mu_2^2 -\xi^2 \right)}} \ ,
\ee
and thus we conclude that
\be	
\braket{\mu_1 ,\mu_2}{\mu_1 , \mu_2}^{\text{c}} =  16\xi ^4 L^2 \left( 1 -\frac{2}{L} \cdot 
\frac{\left( \mu_1^2 +\mu_2^2 -2\xi^2 \right)}{\left[ (\mu_2 -\mu_1)^2 -4\xi^2 \right]} \right) \ . 
\ee
Therefore, at leading order the norm contributes with a factor $\sqrt{L}$ for each magnon and it does not contain any momentum dependence. 
The properly normalised correlation function will be
\begin{align}
	\frac{\granesperado{\lambda}{\sigma^+_k}{\mu_1 \mu_2}^{\text{c}}}{\sqrt{\braket{\l}{\l}^{\text{c}} \braket{\mu_1 ,\mu_2}{\mu_1 , \mu_2}^{\text{c}}}} 
	&=\frac{e^{i(p_1+p_2-p_\lambda )k}}{\sqrt{L^3}} \frac{2}{\mu_2-\mu_1+i} \, 	\left[ \frac{\mu_2^2-\xi^2}{(\mu_2-\lambda)} -\frac{\mu_1^2-\xi^2}{(\mu_1-\lambda)} \right] \notag \\
	&\times \left( 1 -\frac{2}{L} \cdot \frac{\left( \mu_1^2 +\mu_2^2 -2\xi^2 \right)}{\left[ (\mu_2 -\mu_1)^2 -4\xi^2 \right]} \right)^{-1/2} \ .
\end{align}
We want to stress that these expressions for $\granesperado{\lambda}{\sigma^+_k }{\mu_1 \mu_2}$ are \emph{valid to all orders in $L$} provided that we use an expression for the rapidities valid to all orders in $L$.\footnote{As long as the Bethe ansatz is exact and not only an asymptotic approximation.} 
We can thus write the rapidities in terms of the momenta, 
$\mu=- \frac{1}{2} \cot \left( \frac{p}{2} \right)$ and expand in the length of the chain. In the single-magnon state, the momentum is quantised as 
\be
p_{\l} =\frac{2\p n_{\l}}{L} \ .
\ee
In the two-magnon state, the solution to the Bethe equations can be expanded as
\be
p_{1} = \frac{2\pi n_1}{L} + \frac{4\pi}{L^2} \frac{n_1 n_2}{n_1-n_2} + \order{L^{-3}} \ , \quad 
p_{2} = \frac{2\pi n_2}{L} - \frac{4\pi}{L^2} \frac{n_1 n_2}{n_2-n_1}+\order{L^{-3}}  \ .
\ee
We conclude that for the case of $k=1$
\begin{align}
    & \granesperado{\lambda}{\sigma^+_{k=1}}{\mu_1 \mu_2}^{\text{c}} = \frac{1}{\sqrt{L^3}} \frac{2 n_\l (n_1+n_2-n_\l)}{(n_\l -n_1) (n_\l -n_2)}
    \left[ 1 + \frac {1}{L \ (n_1-n_2)^2}  \big[ (n_1^2 +n_2^2) \right. \notag \\
    & \left. + \: \frac{4 n_1^2 n_2^2}{(n_\l-n_1)(n_\l-n_2)} + 2 i \pi (n_1-n_2) ( n_1^2 -n_2^2 +n_1 n_2 - n_\l (n_1-n_2) ) \big] + \cdots \right] \ .
\end{align}

For comparison, let us provide some detail on this computation using the CBA. For that, we need to know the explicit expressions of the coefficients $A(\sigma,\vec{p})$ for one and two magnons. 
If we set $A(\mathbb{I},\vec{p})=1$ we can completely fix the wavefunction for the case of one magnon, while the case of two magnons is given by equation~\eqref{twomagnonwavefunction}. 
Substituting these coefficients in (\ref{esperadoCBA}) and summing over permutations we get
\begin{multline}
    \granesperado{\lambda}{\sigma^+_{k}}{\mu_1 \mu_2}^{\text{c}}= \sum_{1 \leq l<k} \left[e^{i (p_1 - p_\lambda)l + ip_2 k} + S(p_2 , p_1) e^{i (p_2 - p_\lambda)l + ip_1 k} \right] +\\
    + \sum_{k<l \leq L} \left[e^{i p_1 k + i(p_2 - p_\lambda) l} + S(p_2 , p_1) e^{i p_2 k + i(p_1 - p_\lambda) l} \right] \ .
\end{multline}
Although the sums we have to deal with are just regular geometric sums, they do not cover the complete range from $l=1$ to $l=L$, which slightly complicates the computation. 
Let us focus on the case $k=1$, where one of the sums disappears and the other can be easily completed to the full range
\begin{equation}
    \granesperado{\lambda}{\sigma^+_{k=1}}{\mu_1 \mu_2}^{\text{c}}= \sum_{1 \leq l \leq L} \left[e^{i p_1 + i(p_2 - p_\lambda) l} + S(p_2 , p_1) e^{i p_2 + i(p_1 - p_\lambda) l} \right] - \left( 1+ S(p_2 , p_1) \right) e^{i (p_1 p_2 - p_\lambda) } \ .
\end{equation}
If the momenta were all quantised to be $\frac{2\pi n}{L}$, the sum would have provided just some Kronecker deltas. However, $p_1$ and $p_2$ have some additional structure, 
which makes them more involved, and periodicity has to be replaced by the Bethe equations. For example, 
\begin{equation}
    \sum_{1 \leq l \leq L} e^{i (p_2 - p_\lambda) l }=e^{i (p_2 - p_\lambda) }\frac{e^{i (p_2 - p_\lambda) L }-1}{e^{i (p_2 - p_\lambda) }-1}=\left\{\begin{array}{lr}
        L+2\pi i \frac{n_1 n_2}{n_1 - n_2} + \cdots & \text{if } n_\lambda=n_2\\
        \frac{-2n_1 n_2}{(n_\lambda - n_2) (n_1 - n_2)}+ \cdots & \text{if } n_\lambda \neq n_2
        \end{array}\right. \ .
\end{equation}
The full computation can be found in \cite{KM1}. After carefully evaluating the sums, using the Bethe equations and a healthy dose of algebra it is obtained
\begin{equation}
    \granesperado{\lambda}{\sigma^+_{k=1}}{\mu_1 \mu_2}^{\text{c}} = \frac{2 n_\l (n_1+n_2-n_\l)}{(n_\l -n_1) (n_\l -n_2)} \ ,
\end{equation}
if all the mode numbers $n_i$ are different. This expression matches the leading order of the correlation function that we obtained using the ABA, up to the factor $L^{-3/2}$ that can be retrieved by normalising $A(\mathbb{I},\vec{p})=L^{-N/2}$.

It is clear that the computation using the CBA is more involved than the computation using the ABA. The former requires us to perform a total of $(N-1) ((N-1)!)N!$ geometric sums, which we then have to sum and simplify using the Bethe equations. 
The latter one requires us to compute two determinants, one of which can be immediately written as a product. In addition, it already incorporates information about the Bethe equations, simplifying even more the computations. 
As the number of excitations grow, it is clear that the computation using the ABA becomes more and more efficient compared to the one using the CBA.

As a final comment, we can get a more compact result, valid to all orders in $L$, if we take into account the trace condition (\ref{trace}), which in the two-magnon state becomes $\mu_1=-\mu_2$. In this case, the Bethe equations can be solved analytically, 
\footnote{We impose the trace condition on the two-magnon state but not on the single-magnon state because the latter forces the correlation function to vanish.}
\be
\mu_1 = -\mu_2=-\frac{1}{2} \cot \left( \frac{n \, \pi}{L-1} \right) \ , \quad n\in \mathbb{Z} \ . \label{rapidityroiban}
\ee
Substituting this, we obtain
\begin{align}
& \frac{\granesperado{\lambda}{\sigma^+_k}{\mu, -\mu}^{\text{c}}}{\braket{\l}{\l}^{\text{c}} \braket{\mu ,-\mu}{\mu , -\mu}^{\text{c}}} 
= \frac{e^ {-ip_\l k}}{L \sqrt{(L-1)}} \frac{2\m (\m + \xi )}{\m ^2 -\l ^2} \nonumber \\
& = e^ {-2\pi i n_\l k/L}\frac{\cot \left( \frac{n \, \pi}{L-1} \right) }{L \sqrt{(L-1)}} 
\frac{ 2\left[ \cot \left( \frac{n \, \pi}{L-1} \right) -i\right]}{\cot ^2 \left( \frac{n \, \pi}{L-1} \right) -\cot ^2 \left( \frac{n_\l \, \pi}{L} \right) } \ ,
\end{align}
where $n$ and $n_\l$ are integer numbers.

The case of correlation functions for the operator $\sigma^+_k$ evaluated between Bethe states with higher number of magnons can be studied similarly. First, commuting the $(A+D)^n$ factors that surround the $C$ operator is immediate, 
as they are Bethe states. Hence, the only difficult step
is to compute the scalar product between an on-shell state, $\{\mu\}$, and an off-shell state,
$\{\{\lambda\} , \xi\}$, using equation (\ref{howtoscalarproduct}). A detailed computation of these correlation functions can be found in section 3.1.6 of~\cite{Mossel}.

Of course, the computation of correlation functions of the operator $\sigma^-_k$ can be obtained
as the conjugate of the computation we have presented in this section. The case of $\sigma^z$ is a bit more involved, as we have to commute the operator $A-D$ through a stack of $B$ or $C$ operators. 
We will leave it for the next section, as there we will consider the case of the operator $\sigma^+ \sigma^-$. Nevertheless, a detailed computation of these correlation functions can also be found in section 3.1.6 of~\cite{Mossel}.

%%%%%%%%%%%%%%%%%%%%%%%%%%%%%%%%%%%%%%%%%%%%%%%%%%%%%%%%%%%%%%%%%%%%%%%%%
%%%%%%%%%%%%%%%%%%%%%%%%%%%%%%%%%%%%%%%%%%%%%%%%%%%%%%%%%%%%%%%%%%%%%%%%%

\subsection{Correlation functions involving two operators in the ABA}
\label{evaluationcorrelationsection}

In the previous section, we have described how to compute correlation functions of single spin operators using the ABA. Let us move now to the case of two operators, where the difficulty greatly increases.
This happens because the correlation functions we have to compute are of the form
\begin{displaymath}
	\granesperado{0}{\dots C(\xi) (A+D)^n (\xi) B(\xi) \dots}{0} \ .
\end{displaymath}
Therefore, according to the algebra~(\ref{commCAD}), whenever we try to commute the $(A+D)$ operators with the $C$ operator, 
an indeterminate form should appear. In this section, we show that these indeterminate forms yield a finite result, and we explain how to obtain it. We will start by detailing how to proceed 
in the most simple case, that is, when we have only one operator $C$ at the left of the stack of $(A+D)$ factors. As we do not want any further complications that may muddy the calculations, 
we will focus on $\langle 0|\sigma^+_k \sigma^-_k |0\rangle$ and $\langle 0|\sigma^+_k \sigma^+_l |\mu_1 \mu_2\rangle$, both of which are trivial to compute using the CBA.

\subsubsection{First evaluation of $\langle 0|\sigma^+_k \sigma^-_k |0\rangle$}

The correlation function $\langle \{\mu\}|\sigma^+_k \sigma^-_k |\{\lambda\}\rangle$ can be computed in two different ways: we can either consider the operators $\sigma^+_k \sigma^-_k $ separately, 
which leads to a correlation function of the structure we are interested to study, or use the relation $2 \sigma^+ \sigma^- = \mathbb{I} + \sigma^z$, which leads to computation free of divergences. 
Here we will compute it using the second approach and leave the first approach for the next section.

Combining (\ref{sigmaz}) with the fact that the identity operator should correspond to $\prod_{i=1}^L (A+D) (\xi_i)$, we can write 
\begin{equation}
        \sigma^+_k \sigma^-_k= \left(\begin{array}{cc}
        1 & 0 \\
        0 & 0
\end{array} \right)_k= (A+D)^{k-1} (\xi ) A(\xi ) (A+D)^{L-k} (\xi ) \ .
\end{equation}
As in the previous section, the action of the $(A+D)^n$ factors over the Bethe states is known to give the exponential of the momenta,
\begin{equation}
        \granesperado{ \{\mu\}}{\sigma^+_k \sigma^-_k }{\{\lambda\}}=\granesperado{ \{\mu\}}{A(\xi ) }{\{\lambda\}}e^{i\sum_{i}{p_{\mu_i} (k-1)}-i\sum_{i}{p_{\lambda_i} (L-k)}} \ .
\end{equation}
Now we can apply the commutation relation~(\ref{commAB}) to move the $A(\xi)$ operator all the way to the right (similarly, we can do it by moving it to the left). After some effort, we can write
\begin{equation}
        A(\xi) \ket{\{\lambda \}}=\prod_{i=1}^M{\frac{\xi-\lambda_i-i}{\xi-\lambda_i}} \ket{\{\lambda\}} +\sum_{i=1}^M{ \frac{i}{\xi-\lambda_i} \prod_{j\neq i}{\frac{\lambda_i-\lambda_j-i}{\lambda_i-\lambda_j}} \ket{\{\hat{\lambda}_i \}, \xi}} \ ,
\end{equation}
where $\{\hat{\lambda}_i \}$ means that the rapidity $\lambda_i$ is missing from the set $\{\lambda\}$. If we now apply this expression to the bra state $\bra{\{\mu\}}$, the first term of the sum will involve 
the scalar product between both on-shell states, which means that it will only contribute when $\{\lambda\}=\{\mu\}$. The rest of the terms will involve the computation of off-shell-on-shell scalar products. 
The final answer is then
\begin{equation}
        \granesperado{ \{\mu\}}{\sigma^+_k \sigma^-_k }{\{\lambda\}}=\braket{\{\mu\}}{\{\lambda\}} \prod_{i=1}^M{\frac{\xi-\lambda_i-i}{\xi-\lambda_i}} +\sum_{i=1}^M{ \frac{i\braket{\{\mu\}}{\{\hat{\lambda}_i \}, \xi}}{\xi-\lambda_i} \prod_{j\neq i}{\frac{\lambda_i-\lambda_j-i}{\lambda_i-\lambda_j}}} \ .
\end{equation}

To get a better understanding of this formula and some subtleties that arise, we are going to compute explicitly the cases with a few magnons. Obviously, the most trivial case is $\{\lambda\}=\{\mu\}=\emptyset$, which reads
\begin{equation}
        \granesperado{0}{\sigma^+_k \sigma^-_k }{0}=\granesperado{0}{A(\xi ) }{0}=a(\xi )=1 \ .
\end{equation}
Which trivially agrees with the result we get using CBA. The first non-trivial case is the one with a single magnon, which reads
\begin{equation}
        \granesperado{ \mu}{\sigma^+_k \sigma^-_k }{\lambda}=\left\{ \begin{array}{cc}
        \frac{-e^{i(p_\lambda - p_\mu) k} }{(\lambda + \xi ) (\mu-\xi )} &\text{ if } \lambda \neq \mu\\
        \frac{L}{(\lambda + \xi ) (\mu-\xi )} \left( 1-\frac{1}{L} \right) &\text{ if } \lambda = \mu
\end{array}       \right. \ .%}                                                                                                                                                                             
\end{equation}
The second one is easily comparable with the CBA, because the result can be expressed as $\granesperado{ \lambda}{\sigma^+_k \sigma^-_k }{\lambda}=\braket{\lambda}{\lambda} \left( 1-\frac{1}{L} \right)$, 
which agrees with the CBA. In the same way, after properly normalizing, we obtain that the case $\lambda \neq \mu$ is equal to $e^{i(p_\lambda - p_\mu) (k-\frac{1}{2})}$, which is exactly what we obtain from the CBA shifted by half of a lattice space, as we discussed.

Although these two cases have no problems, the case $\granesperado{ \lambda_1 \lambda_2}{\sigma^+_k \sigma^-_k }{\lambda_1 \lambda_2}$  has to be treated with care. In particular, intermediate steps 
involve computing the correlation functions $\braket{ \lambda_1 \xi}{\lambda_1 \lambda_2}$ and $\braket{ \xi \lambda_2}{\lambda_1 \lambda_2}$, which apparently diverge when we naively substitute the rapidities 
on the Slavnov determinant (\ref{scalarproduct}). This apparent divergence appears because this scalar product have already been simplified using the Bethe equations. Therefore, to find the correct answer 
we have to compute first $\braket{ \mu_1 \mu_2}{\lambda_1 \lambda_2}$ considering both $\{\mu\}$ and $\{\lambda\}$ as off-shell rapidities and, without imposing the Bethe equations, 
compute either limit $\mu_i \longrightarrow \{ \lambda_1 , \xi \}$ or $\mu_i \longrightarrow \{ \xi , \lambda_2 \}$ respectively. \footnote{We want to thank N. A. Slavnov for discussions about this subject.}

\subsubsection{Second evaluation of $\langle 0|\sigma^+_k \sigma^-_k |0\rangle$}

Let us now compute again the correlation function $\langle 0|\sigma^+_k \sigma^-_k |0\rangle$ from the other perspective. This section may seem a seriously overcomplicated way of calculating a trivial correlation function, 
but that is its main point: to show the complexity of this computation without any other interference.

The starting point is again the relations between local spin operators in the CBA and the elements of the monodromy matrix (\ref{sigmaplus}) and (\ref{sigmaminus}). If we recall that $(A+D) (\xi_i) \ket{0}=\ket{0}$ 
for the Heisenberg chain, the correlation function becomes
\begin{equation}
        \langle 0 | \sigma^+_k \sigma^-_l |0\rangle =
        \langle 0| {C (\xi) (A+D)^{L+l-k-1} (\xi) B (\xi)} |0 \rangle \ .
\end{equation}
In order to evaluate this correlation function, we have to commute the operators $(A+D)$ through either the $B$ or $C$ operators using equation (\ref{commBAD}) or (\ref{commCAD}) respectively.
However, although it seems that when trying to commute $(A+D)^n$ we should obtain a pole of order $n$ because the commutation relations seem divergent when the two rapidities are equal, the residue turns out to be zero and the expression is thus finite.
In order to show this cancellation, some care will be needed. Let us first introduce
some notation. We will define
\begin{align}
{\cal F}^L_{n}(\alpha,\delta) &= \granesperado{0}{C(\xi+\alpha) \mathcal{O}(\delta)}{0} \ , \nonumber \\
{\cal F}^L_{n+1}(\alpha,\delta) &= \lim_{\beta\rightarrow \alpha} 
\granesperado{0}{C(\xi+\alpha) (A+D)(\xi+\beta) \mathcal{O}(\delta)}{0}= \lim_{\beta\rightarrow \alpha} f^L_{n+1}(\a , \b , \d )\ , 
\label{F}
\end{align}
where $\mathcal{O}(\delta)$ denotes a generic operator. The reason for the subindex $n$ is that in all the cases that we will consider $\mathcal{O}(\delta)$ 
will include a factor $(A+D)^n$. Now, using (\ref{commCAD}) we can write
\begin{align}
	{\cal F}^L_{n+1} (\alpha,\delta)&=\left[ 1+d(\xi+\alpha) \right] {\cal F}^L_{n}(\alpha,\delta) 
	+ \lim_{\b \rightarrow \a}\frac{i}{\b-\a} \left\{ \left[d(\xi+\b)-1 \right] {\cal F}^L_{n}(\alpha,\delta) \right. \nonumber \\
	&-\left. \left[ d(\xi+\alpha)-1 \right] {\cal F}^L_{n}(\b,\delta) \right\} \ ,
\label{Fn1}
\end{align}
If we expand in a Taylor series, we find that all terms of order $1/(\b-\a)$ cancel themselves. 
Therefore, we can safely take the limit $\b \rightarrow \a$ to get 
\begin{equation}
	{\cal F}^L_{n+1} (\alpha,\delta)=\left[1+d(\xi+\alpha)+i\left.\dpartial{d}{\lambda} \right|_{\xi+\alpha} \right] {\cal F}^L_n (\alpha, \delta) + 
	i \big[ 1-d(\xi+\alpha) \big] \dpartial{{\cal F}^L_n (\alpha, \delta)}{\alpha} \ .
\label{Frecurrence}
\end{equation}
We should stress that the derivative with respect to $\a$ in this expression must be understood with respect to the argument of the $C$ operator. 
As a consequence, it does not act on the rest of the operators. This will introduce some subtleties in the next step of the calculation.
The idea now is to use (\ref{Frecurrence}) as a recurrence equation to find $\langle 0|\sigma^+_k \sigma^-_k |0\rangle$.
However, this is not straightforward, because it requires information on correlation functions of the form 
\be
\granesperado{0}{C(\xi+\alpha) (A+D) (\xi+\delta) \dots}{0} \ ,
\label{fn}
\ee
but returns instead information about correlators of the form 
\be
\granesperado{0}{C(\xi+\alpha) (A+D) (\xi+\alpha) (A+D) (\xi+\delta) \dots}{0} \ ,
\label{fn1}
\ee
where the argument of the first $(A+D)$ factor in (\ref{fn1}) also depends on $\a$. Thus, 
in order to obtain the expression we actually want, we should take the derivative with respect to $\a$ in $f^L_{n+1} (\a ,\b ,\d)$, and then take the limit $\b \rightarrow \a$, 
instead of taking directly the derivative in ${\cal F}^L_{n+1}(\alpha,\delta)$.
Therefore, using~(\ref{Fn1}),
\begin{align}
	\lim_{\b \rightarrow \a} \dpartial{ f^L_{n+1} (\alpha,\b ,\delta)}{\alpha} & = 
	\big[ 1+d(\xi+\alpha) \big] \dpartial{{\cal F}^L_{n}(\alpha,\delta)}{\alpha} + \lim_{\b \rightarrow \a}\frac{i}{\b-\a} 
	\left\{ \big[d(\xi+\b)-1 \big] \dpartial{{\cal F}^L_{n}(\alpha,\delta)}{\alpha} \right. \nonumber \\
	& - \dpartial{d(\xi+\alpha)}{\alpha} {\cal F}^L_{n}(\b,\delta) 
	+ \frac {1}{(\b-\a )^2} \Big[ \big[d(\xi+\b)-1 \big] {\cal F}^L_{n}(\alpha,\delta) \nonumber \\
	& \left. - \big[d(\xi+\alpha)-1 \big] {\cal F}^L_{n}(\b,\delta) \Big] \right\} \ .
\end{align}
The limit is similar 
to the previous one. In this case, after a series expansion we find a pole of order two and a pole of order one, but they cancel each other. 
The final result is
\begin{align}
	\lim_{\b \rightarrow \a} \dpartial{ f^L_{n+1} (\alpha,\b ,\delta)}{\alpha}&=\big[ 1+d(\xi+\alpha) \big] 
	\dpartial{{\cal F}^L_{n} (\alpha,\delta)}{\alpha}+\frac{i}{2} \dpartial[2]{d}{\alpha} {\cal F}^L_{n} (\alpha,\delta) \nonumber \\
	&+ \frac{i}{2} \big[ 1-d(\xi+\alpha) \big] \dpartial[2]{{\cal F}^L_{n} (\alpha,\delta)}{\alpha} \ .
\end{align}
So far we have proved that when we have one derivative and we commute one $(A+D)$ factor we get another 
derivative over the correlation function. In general, if we have $m$ derivatives we get
\begin{align}
	\lim_{\b \rightarrow \a} \dpartial[m]{ f^L_{n+1} (\alpha,\b ,\delta)}{\alpha}& = 
	\big[ 1+d(\xi+\alpha) \big] \dpartial[m]{{\cal F}^L_{n} (\alpha,\delta)}{\alpha}+\frac{i}{m+1} 
	\dpartial[m+1]{d}{\alpha} {\cal F}^L_{n} (\alpha,\delta) \nonumber \\
	&+\frac{i}{m+1} \big[ 1-d(\xi+\alpha) \big] \dpartial[m+1]{{\cal F}^L_{n} (\alpha,\delta)}{\alpha} \ ,
\label{Fgeneral}
\end{align}
this identity can be easily proved if we assume that the left-hand side of the equation has no poles. 
Under this assumption, we only need to track the terms without a $\b-\a$ factor when we expand in a Taylor series,
\begin{align}
	& \lim_{\b \rightarrow \a} \dpartial[m]{ f^L_{n+1} (\alpha,\b ,\delta)}{\alpha}= \big[1+d(\xi+\alpha) \big] 
	\dpartial[m]{{\cal F}^L_n (\a , \d)}{\alpha} \nonumber \\ 
	&+ \lim_{\b \rightarrow \a} \, \dpartial[m]{}{\alpha}
	\left\{ \frac{i}{\b-\a} \left[ \Big( d(\xi+\b)-1\Big) {\cal F}^L_{n}(\alpha,\delta) \right. \right.
	-\left. \left. \Big(d(\xi+\alpha)-1\Big) {\cal F}^L_{n}(\b,\delta) \right] \right\} \ .
\label{Fprove}	
\end{align}
The second term on the right-hand side of this expression can be written as
\[
\lim_{\b \rightarrow \a} \sum_j{\binom{m}{j} \frac{i}{(\b -\a)^{j+1} \, (j+1)} } \cdot 
\left[ \dpartial[j+1]{d}{\alpha} \dpartial[m-j]{{\cal F}^L_{n}}{\alpha}  
- \dpartial[m-j]{(d-1)}{\alpha} \dpartial[j+1]{{\cal F}^L_{n}}{\alpha} \right] (\b -\a)^{j+1} + \cdots \ ,
\]
where the dots stand for terms proportional to $(\b-\a)^k$. It is clear that the addends are antisymmetric under the exchange $j\rightarrow m-j-1$, 
allowing us to cancel all the terms except the term $j=m$, which does not have a partner. Substituting it in (\ref{Fprove}), we recover equation (\ref{Fgeneral}).

Let us summarise our results up to this point. We have obtained a complete set of recurrence equations
\begin{align}
	{\cal F}^L_{n+1}(\alpha)&= \left[1+d(\xi+\alpha)+i\left.\dpartial{d}{\lambda} \right|_{\xi+\alpha} \right] {\cal F}^L_n (\alpha) 
	+ i\big[ 1-d(\xi+\alpha) \big] \mathcal{D} {\cal F}^L_n (\alpha) \ , \nonumber \\
	\mathcal{D}^m {\cal F}^L_{n+1} (\alpha)&=\big[ 1+d(\xi+\alpha) \big] \mathcal{D}^m {\cal F}^L_{n} (\alpha) 
	+ \frac{i}{m+1} \dpartial[m+1]{d}{\alpha} {\cal F}^L_{n} (\alpha) \nonumber \\
	&+\frac{i}{m+1} \big[ 1-d(\xi+\alpha) \big] \mathcal{D}^{m+1} {\cal F}^L_{n} (\alpha) \ , \nonumber \\
	\mathcal{D}^m {\cal F}^L_{0} (\alpha) &=\dpartial[m]{{\cal F}^L_{0} (\alpha)}{\alpha} \ , \quad \hbox{with} \quad
	{\cal F} (\alpha )=\lim_{\delta\rightarrow \alpha}{\cal F} (\alpha , \delta) \ , 
\label{recurrenceeq}
\end{align}
where $\mathcal{D}$ is just a convenient notation to refer both to the derivative and the limit, 
\be 
\mathcal{D}^m {\cal F} (\alpha )=\lim_{\substack{\delta\rightarrow \alpha \\ \b \rightarrow \a}} \dpartial[m]{ f(\alpha , \b ,\delta)}{\alpha} \ .
\ee 

Now we are ready to calculate the correlation function provided a starting condition is given. In our case, using~(\ref{commCB}),
\be
{\cal F}^L_0(\alpha)=\granesperado{0}{C(\xi) B(\xi+\alpha)}{0}=-\frac{ic}{\alpha} \frac{\alpha^L}{(\alpha +i)^L} \ ,
\ee
which takes values ${\cal F}_0^1 (0)=-c=1$ and ${\cal F}^{L>1}_0=0$. In order to find $\langle 0|\sigma^+_k \sigma^-_k |0\rangle$ 
we have to calculate ${\cal F}_{L-1}^L (0)$. Because ${\cal F}^L_0(0)$ has a zero of order $L-1$, the only terms that can contribute are those which 
involve a number of derivatives of ${\cal F}^L_0(\alpha)$ greater than or equal to $L-1$.
In appendix~\ref{A} we will construct the correlation function ${\cal F}^L_n(\alpha)$ in full generality, but in this case it is easy to see that
\begin{equation}
	{\cal F}_{L-1}^L (\alpha)=\frac{i^{L-1}}{(L-1)!} \dpartial[L-1]{{\cal F}^L_0(\alpha)}{\alpha}+\dots=\frac{i^{L-1}}{(L-1)!} \cdot i \, 
	\frac{(L-1)!}{i^L}+ \order{\alpha} \ .
\end{equation}
From here, we conclude that the value of this correlator is one, as expected from the CBA. 
We can also prove that ${\cal F}_{n}^L (\alpha)=0$ for $0\leq n <L-1$, which also agrees with the result 
$\langle 0|\sigma^+_k \sigma^-_l |0\rangle=0$ when $k\neq l$ of the CBA.

Let us finish with some words about the case where the  bra state is not empty. We may think that the difficulty does not increase too much, as the operators $A+D$ commute easily 
with a stack of $C$ operators whose rapidities satisfy the Bethe equations. However, we have to keep in mind that the $A+D$ operator has to pass first through $C(\xi + \a)$, 
which add a term with $A-D$ instead, making the computation much involved. As an example, we have collected the case where the bra state is a single-magnon state 
in appendix~\ref{C}. \footnote{We may think that we can use crossing symmetry to empty the bra state and simplify our computations, as it is usually done in the context of form factors, see~\cite{KM1}. Sadly, this is not possible here, as we cannot access crossing. We can get a nicer picture of this from the AdS/CFT correspondence. There, the long-range spin chain has a well-defined crossing transformation, but the Heisenberg spin chain corresponds to the limit where one of the periods of the torus that uniformised the magnon dispersion relation becomes infinitely large, thus forbidding us the access to the crossing transformation.}

%%%%%%%%%%%%%%%%%%%%%%%%%%%%%%%%%%%%%%%%%%%%%%%%%%%%%%%%%%%%%%%%%%
%%%%%%%%%%%%%%%%%%%%%%%%%%%%%%%%%%%%%%%%%%%%%%%%%%%%%%%%%%%%%%%%%%

\subsubsection{Evaluation of $\langle 0|\sigma^+_k \sigma^+_l |\mu_1 \mu_2\rangle$}
\label{0sigmasigmamumu}

Let us move now to the correlation function $\langle 0|\sigma^+_k \sigma^+_l |\mu_1 \mu_2\rangle$.
Using relation (\ref{sigmaplus}), we can again write the correlation function only in terms of elements of the monodromy matrix, 
\be
\langle 0|\sigma^+_k \sigma^+_l |\mu_1 \mu_2\rangle 
= \granesperado{0}{(A+D)^{k-1} (\xi) C(\xi) (A+D)^{n} (\xi) C(\xi) (A+D)^{L-l} (\xi) }{\mu_1 \mu_2} \ ,
\label{wavefunction}
\ee
where $n=L+l-k-1$. The first factor $(A+D)$ acts trivially on the vacuum. On the contrary, the last factor $(A+D)$ 
acts on the two magnon state $\ket{\mu_1 \mu_2}=B(\mu_1) \, B(\mu_2) \, \ket{0}$ and provides a factor $e^{-i(p_1+p_2)\cdot (L-l)}=e^{i(p_1+p_2)l}$, 
where in the last equality we have used the periodicity condition for the Bethe roots. Let us denote the remaining correlation function with $n$ inner factors of $(A+D)$ 
between the $C$ operators by ${\cal G}^L_n(\alpha)$,
\be
{\cal G}^L_{n}(\a) = \granesperado{0}{C(\xi+\alpha) (\mathcal{A+D})^n (\xi+\delta) C(\xi) B(\mu_1) B(\mu_2)}{0} \ .
\ee
Comparing with (\ref{Fn1}), the ${\cal G}^L_{n}(\a)$ functions have the same structure as the ${\cal F}^L_{n}(\a)$. This is good news, as it means that they have 
to fulfil the same recursion relations (\ref{recurrenceeq}), with the only difference being the starting condition, i.e.,
\be
{\cal G}^L_0(\alpha)=\granesperado{0}{C(\xi+\alpha) \, C(\xi) \, B(\mu_1) \, B(\mu_2)}{0}=\langle 0|\sigma^+_1 \sigma^+_{L} |\mu_1 \mu_2\rangle \ .
\ee
This is the product of an on-shell Bethe state with an off-shell Bethe state, and can be computed using the tools described in section~\ref{Bethe}
\be
 \granesperado{0}{C(\xi+\alpha) \, C(\xi) \, B(\mu_1) \, B(\mu_2)}{0} = \frac{\det T}{V} \ ,
\ee
where the matrices $T$ and $V$ are
\begin{align}
	T_{11} &= \frac{-2\xi}{(\mu_1-\xi)^2} \frac{\mu_2+\xi}{\mu_2-\xi} \ , \quad 
	T_{21} =\dpartial{\tau (\xi,\{\mu_1 , \mu_2\} )}{\mu_2} = \frac{\mu_1+\xi}{\mu_1-\xi} \frac{-2\xi}{(\mu_2-\xi)^2} \ , \nonumber \\
	T_{12} &=\frac{-2\xi}{(\mu_1-\xi-\alpha)^2} \, \frac{\mu_2+\xi-\alpha}{\mu_2-\xi-\alpha}+\frac{\alpha^L}{(i+\alpha)^L} \, 
	\frac{2\xi}{(\mu_1-\xi-\alpha)^2} \, \frac{\mu_2-3\xi-\alpha}{\mu_2-\xi-\alpha} \ , \nonumber \\
	T_{22} &= \frac{\mu_1+\xi-\alpha}{\mu_1-\xi-\alpha} \, \frac{-2\xi}{(\mu_2-\xi-\alpha)^2}+\frac{\alpha^L}{(i+\alpha)^L} \, 
	\frac{\mu_1-3\xi-\alpha}{\mu_1-\xi-\alpha} \, \frac{2\xi}{(\mu_2-\xi-\alpha)^2} \ , \nonumber \\
	\frac{1}{V} &=\frac{(\mu_1 -\xi) (\mu_1 -\xi-\alpha)(\mu_2 -\xi)(\mu_2 -\xi-\alpha)}{\alpha (\mu_1 -\mu_2)} \ .
\end{align}
After some algebra we can easily organise ${\cal G}_0^L(\a)$ as an expansion in $\a$,
\ba
{\cal G}^L_0 (\alpha) & \!\!\! = \!\!\! & \left( A_0+\alpha A_1+\alpha^2 A_2 + \cdots \right)+\alpha^{L-1} 
\left( B_{L-1}+\alpha B_{L}+\alpha^2 B_{L+1} + \cdots \right) \nonumber \\ 
& + \!\! & \alpha^{2L-1} \left( C_{2L-1} + \a C_{2L} + \a^2 C_{2L+1} + \cdots \right) \ ,
\label{FLzero}	
\ea
with $A_q$ and $B_{L+q-1}$ given by \footnote{Thanks to the periodicity condition, we will not need the explicit expression for $C$.}
\ba
A_q & \!\! = \!\! & \frac{1}{\mu_1-\mu_2} \, \frac{\mu_1^+ \mu_2^+}{\mu_1^- \mu_2^-} \, 
\left[ \frac{1}{(\mu_1^-)^{q}}    \frac{\left(\mu_2-\mu_1+i \right)}{\mu_1^- \mu_2^+} 
+ \frac{1}{(\mu_2^-)^{q}} \frac{\left(\mu_2-\mu_1-i \right)}{\mu_1^+ \mu_2^-} \right] \ , \nonumber \\
B_{L+q-1} & \!\! = \!\! & \sum_{j=0}^{q}{i^j \binom{L+j-1}{j} \beta_{q-j}} \ , 
\label{AB}
\ea
where we have defined
\ba
	\beta_0 & \!\! = \!\! & B_{L-1}=\frac{1}{i^L} \, \frac{1}{\mu_1^- \mu_2^-}\, \frac{1}{\mu_1-\mu_2} \, 
	\left( \mu_2^+ \mu_1^{---} - \mu_1^+ \mu_2^{---} \right) \ , \nonumber \\
	\beta_q & \!\! = \!\! & \frac{1}{i^L} \, \frac{1}{\mu_1^- \mu_2^-} \, \frac{1}{\mu_1-\mu_2} \, 
	\left( \frac{\mu_2^+ \mu_1^{---} -\mu_2^+ \mu_2^-}{(\mu_2^-)^q} - \frac{\mu_1^+ \mu_2^{---} -\mu_1^+ \mu_1^-}{(\mu_1^-)^q} \right) \ , 
\ea
with $\mu_i^k=\mu_i+k\xi$ and $B_q=C_p=0$ for $q<L-1$ and $p<2L-1$.
The next step is to find the general form of the correlation function ${\cal G}^L_n(\a)$. Using the recurrence equations~(\ref{recurrenceeq}) 
the first terms can be easily calculated for a general value of $\a$,
\begin{align}
	{\cal G}^L_1(\a) &=\left[ 1+d+i \dpartial{d}{\lambda}  \right] {\cal G}^L_0(\a)+i \big[ 1-d \big] \dpartial{{\cal G}^L_0(\a)}{\lambda} \ , \nonumber \\
	{\cal G}^L_2(\a) &=\left[ 1+2d+2i\dpartial{d}{\lambda}+2id\dpartial{d}{\lambda}+d^2-\left( \dpartial{d}{\lambda} \right)^2 
	- \frac{1}{2} \dpartial[2]{d}{\lambda}+\frac{d}{2} \dpartial[2]{d}{\lambda} \right] {\cal G}^L_0(\a) \nonumber \\
	& + \left[ 2i-2id^2 -\dpartial{d}{\lambda}+d\dpartial{d}{\lambda} \right] \dpartial{{\cal G}^L_0(\a)}{\lambda} 
	-\frac{(1-d)^2}{2} \dpartial[2]{{\cal G}^L_0(\a)}{\lambda} \ ,
\end{align}
where $d=d(\xi +\alpha)$ and $\dpartial{d}{\lambda}=\left. \dpartial{d}{\lambda} \right|_{\xi +\alpha}$. 
If we take now the limit $\alpha\rightarrow 0$, all the $d$'s and derivatives of $d$ disappear, unless 
the order of the derivative is larger or equal to $L$.

As they follow the same recursion relation, we can borrow our results for ${\cal F}^L_{n}(\a)$ from appendix~\ref{A} and write
\begin{equation}
{\cal G}^L_n(0)=\sum_{q=0}^{n}{\binom{n}{q} \left. \frac{i^{q} \mathcal{D}^{q}}{q!} {\cal G}^L_0(\a) \right|_{\a =0}}+\theta(n-L) {\cal G}^L_{n-L}(0) \ , \label{GLn}
\end{equation}
where $\theta (x)$ is the Heaviside step function. If we use now expansion (\ref{FLzero}) 
and perform the corresponding derivatives, we get
\be
	{\cal G}^L_n(0) = \sum_{q=0}^{n}{\binom{n}{q} i^{q} \left( A_q+B_q+C_q \right)}+\theta(n-L) {\cal G}^L_{n-L}(0) \ .
\label{Ffinal}
\ee
As such, we have to study separately the cases of $n<L-1$ and $n=L$. The case of $n>L-1$ should give us the same results as $n<L-1$ due to periodicity, but we will check it anyway.

\paragraph*{The case $n<L-1$}

We will first consider the case where $n<L-1$, which corresponds to $l<k$. 
From (\ref{Ffinal}) it is clear that when $n<L-1$ the only contribution comes from the $A_q$ coefficients, that can be easily summed up,
\be
\sum_{q=0}^{n}{\binom{n}{q} i^{q} A_q} = \frac{1}{\mu_1-\mu_2} \frac{\mu_1^+ \mu_2^+}{\mu_1^- \mu_2^-} \! 
\left[  \left( \frac{\mu_1^+}{\mu_1^-} \right)^{n} 
\frac{\left(\mu_2-\mu_1+i \right)}{\mu_1^- \mu_2^+} 
+ \left( \frac{\mu_2^+}{\mu_2^-} \right)^{n} \frac{\left(\mu_2-\mu_1-i \right)}{\mu_1^+ \mu_2^-} \right] .
\label{Asum}
\ee
Recalling now that the rapidities parameterise the momenta, $\mu_i^+/\mu_i^-=e^{-ip_i}$, 
the correlation function~(\ref{wavefunction}) can be written as
\be
\granesperado{0}{\sigma^+_k \sigma^+_l}{\mu_1 \mu_2} = 
\frac{1}{\mu_1-\mu_2} \, \frac{\mu_2-\mu_1+i}{\mu_1^- \mu_2^-} \left[   e^{ip_1(k-L)+ip_2 l}  +e^{ip_2(k-L)+i p_1 l} S_{21} \right] \ ,
\ee
where we have inserted the S-matrix, 
\be
S_{21}=\frac{\mu_2-\mu_1-i}{\mu_2-\mu_1+i} \ ,
\ee
and we have taken into account that $n=L+l-k-1$. Using now the Bethe equations 
$e^{-ip_1 L}=e^{ip_2 L}=S$, we find 
\begin{equation}
	\granesperado{0}{\sigma^+_k \sigma^+_l}{\mu_1 \mu_2}=\frac{1}{\mu_1-\mu_2} \, \frac{\mu_2-\mu_1+i}{\mu_1^- \mu_2^-} 
	\left[   e^{i (p_1 k+p_2 l)} S_{21}  +e^{i(p_2 k+ p_1 l)} \right] \ . \label{y<x}
\end{equation}
Note that, although this result is only true as long as $l<k$, we can observe a structure reminiscent of the CBA wavefunction up to the factor in front of the bracket. At the end of this section, we will see how the normalization proposed 
in section~\ref{normalizationsection} allows us to get rid of it.

\paragraph*{The case $n=L-1$}

Our next step is the calculation of ${\cal G}^L_{L-1}(0)$, which must be identically zero, because it corresponds to the case 
where both operators are located at the same site, $k=l$. 
Equation~(\ref{Ffinal}) takes the form\be
{\cal G}^L_{L-1}(0)=i^{L-1} B_{L-1}+\sum_{q=0}^{L-1}{\binom{n}{q} i^{q} A_q} \ .
\ee
The second term is already known from the previous calculation. Therefore, we only need to substitute $n=L$ in (\ref{Asum}) and make use of the Bethe equations
\be
\sum_{q=0}^{L-1}{\binom{L-1}{q} i^{q} A_q} = - \frac{2}{\mu_1^- \mu_2^-} \ . 
\ee
On the other hand
\be
i^{L-1} B_{L-1}=\frac{i}{\mu_1 -\mu_2} \frac{1}{\mu_1^- \mu_2^-} 
\left( \mu_1^+ \mu_2^{---} -\mu_1^{---} \mu_2^+ \right) =
\frac{2}{\mu_1^- \mu_2^-} \ .
\ee
Therefore ${\cal G}^L_{L-1}(0)=0$, as we expected from the CBA.

\paragraph*{The case $n>L-1$}

The last correlation functions that we will evaluate will be those with $L-1<n<2L-1$. Obviously, 
we expect that ${\cal G}^L_{n+L}(0)$ should equal ${\cal G}^L_n(0)$ due to periodicity. In order to prove this statement, we will first show  
that the contribution from the $B$ terms is going to be $\binom{n-L}{q+1} i^{L+q} \beta_{q}$. Next, we will  
show that this coefficient cancels $\sum_{q=0}^{n}{\binom{n}{q} i^{q} A_q}$, and thus we will conclude that ${\cal G}^L_{n+L}(0)={\cal G}^L_n(0)$.
Let us see how it goes. Recalling the expression for $B_{q}$ in (\ref{AB}) and performing the sum, we find
\ba
\sum_{q=L-1}^n{\binom{n}{q} i^q B_q} 
& \!\! = \!\! & \sum_{s=0}^{n-L+1}{\sum_{r=0}^{s}{\binom{n}{s+L-1} \binom{L+r-1}{r} i^{s+r+L-1} \beta_{s-r}}} \ .
\ea
As we would like to extract the factor associated to a particular coefficient $\beta$, we have to change the sum over $t$ by a sum over $q=s-t$ and swap the order of the sum over $s$ and $q$. After doing so, the factor associated to $\b_{q}$ is
 \be
i^{L+q-1} \sum_{r=0}^{n-L-q+1}{\binom{n}{L+r+q-1} \binom{L+r-1}{r} (-1)^r } \ ,
\ee
where we have transformed the sum over $s$ into a sum over $r=s-q$. This is convenient because all terms with $s<q$ do not contribute to $\beta_q$. The sum can be compactly written in terms of a hypergeometric function
\begin{align}
	& \frac{n!}{(L-1)!} \sum_{r=0}^{n-L-q+1}{\frac{(L+r-1)!}{(L+r+q-1)!} \binom{n-L-q+1}{r}  (-1)^r } \nonumber \\
	& = \, _2 F_1 \left( L,q-1+L-n;L+q;1 \right) n! = \frac {(n-L)!}{(q-1)!} \ , 
\end{align}
where in the last equality we have used Kummer's first formula,
\be
_2 F_1 \left( \frac{1}{2}+m-q,-n;2m+1;1 \right)=\frac{\Gamma (2m+1) 
\Gamma \left( m+\frac{1}{2}+q+n \right)}{\Gamma \left( m+\frac{1}{2} +q \right) \Gamma (2m+1+n)} \ .
\ee
Therefore, there is no contribution from $\beta_0$. On the contrary, the rest of the coefficients will contribute with $\binom{n-L}{q-1} i^{L+q-1}$. 
Now if we use that 
\be
\sum_\alpha{\binom{K-L}{\alpha} \frac{i^\alpha}{\left( \mu^- \right)^\alpha}}=\left( \frac{\mu^+}{\mu^-} \right)^{K-L} \ ,
\ee
together with $\mu_1^+ \mu_2^{---}-\mu_1^+ \mu_1^-=\mu_1^+ \left( \mu_2 -\mu_1 -i \right)$, we find that 
\ba
&\sum_{q=0}^{n-L}{\binom{n-L}{q} i^{L+q} \beta_q} \nonumber \\
&= \frac{1}{\mu_1^- \mu_2^-} \, \frac{-1}{\mu_1-\mu_2} \, 
\left[ \left( \frac{\mu_1^+}{\mu_1^-} \right)^{n-L+1} \left( \mu_2 -\mu_1 -i \right) 
+ \left( \frac{\mu_2^+}{\mu_2^-} \right)^{n-L+1} \left( \mu_2 -\mu_1 +i \right) \right] \ . 
\label{sumbeta}
\ea
If we now remove the $-L$ factor by extracting an S-matrix, expression (\ref{sumbeta}) 
cancels exactly the contribution from the sum of the $A$'s in (\ref{Asum}). Finally, we conclude that
\ba
\granesperado{0}{\sigma^+_k \sigma^+_l}{\mu_1 \mu_2} & \!\! = \!\! & \frac{e^{i(p_1+p_2)l}}{\mu_1-\mu_2} \, \frac{1}{\mu_1^- \mu_2^-} 
\left[ e^{ip_1(k-l)}  \left(\mu_2-\mu_1+i \right) +e^{ip_2(k-l)} \left(\mu_2-\mu_1-i \right) \right] \notag \\
& = \!\! & \frac{1}{\mu_1-\mu_2} \, \frac{\mu_2-\mu_1+i}{\mu_1^- \mu_2^-} \left[   e^{i(p_1k+p_2 l)}  +e^{i(p_2k+ p_1 l)} \, S_{21} \right] 
\label{y>x} \ ,
\ea
which agrees with (\ref{y<x}), but with $k$ and $l$ exchanged because now we are in the case $k<l$.

As in the previous section, the resulting expressions highly simplifies if we impose the trace condition (\ref{trace}). 
When we replace the rapidities from equation~(\ref{rapidityroiban}) in these expressions, after some immediate algebra we obtain 
\be 
L \frac{\granesperado{0}{\sigma^+_k \sigma^+_l}{\mu, -\mu}}{\sqrt{\braket{\mu ,-\mu}{\mu , -\mu}}}= 2 \sqrt{\frac{L}{L-1}} \cos \left( \frac{(2|l-k|-1)\pi n}{L-1} \right) \ ,
\label{normalizedtwomagnon}
\ee
with $|l-k|\leq L-1$. This result match the analysis in reference~\cite{RV}, where this correlation function was calculated for the cases 
$l-k=1$ and $l-k=2$  (we have written the factor $L$ on the left-hand-side of (\ref{normalizedtwomagnon}) to follow conventions in there).

%%%%%%%%%%%%%%%%%%%%%%%%%%%%%%%%%%%%%%%%%%%%%%%%%%%%%%%%%%%%%%%%%%
%%%%%%%%%%%%%%%%%%%%%%%%%%%%%%%%%%%%%%%%%%%%%%%%%%%%%%%%%%%%%%%%%%

\section{The long-range Bethe ansatz} \label{longcorr}

In this section, we will repeat the computation we have done in the previous sections, but in the context of the long-range BDS spin chain~\cite{BDS}. 
This can be done quite easily because in all our previous expressions we have kept general the homogeneous point. Therefore, as 
the BDS spin chain can be mapped into an inhomogeneous short-range spin chain, with the inhomogeneities located at
\be 
\xi_n=\frac{i}{2} +\sqrt{2} g\cos\frac{(2n-1)\pi}{2L} \equiv \xi + g \k_n \ ,
\ee
it is rather simple to extend all computations above to the long-range Bethe ansatz. An immediate example is the normalization factor 
for the operator $B(\lambda)$, which is straightforward to calculate given the expressions from section~\ref{normalizationsection},
\be
B(\lambda)=\sum_{n=1}^L{\frac{iS^-_n}{\lambda-\xi_n} \left( \prod_{l=1}^n{ \frac{\lambda-\xi_l}{\lambda-\xi_l+i}} \right) } \ .
\ee
We conclude therefore that in the long-range Bethe ansatz, the difference in normalization between the ABA and the CBA depends 
on the site where the spin operator acts. An analogous result follows for the operator $C(\lambda)$.

Another computation that we can readily extend to the long-range Bethe ansatz is the calculation of scalar products. This is immediate because the solution 
to the inverse scattering problem in expressions~(\ref{sigmaplus})-(\ref{sigmaz}) is valid for an inhomogeneous spin chain. 
Furthermore, equations~(\ref{scalarproduct}) and (\ref{howtoscalarproduct}) can be directly used without modifications. Let us look, for example, at the correlation function of the single-magnon state, 
\begin{equation}
\granesperado{0}{\sigma^+_k}{\lambda}=\frac{i}{\l-\xi-g \k_k} \prod_{j=1}^k{\frac{\l - \xi-g \k_j}{\l+\xi - g \k_j}} \ ,
\end{equation}
which as in the case of the homogeneous spin chain should also be divided by the norm
\begin{equation}
\sqrt{\braket{\l}{\l}}=\sqrt{i \dpartial{d}{\l}} = i \sqrt{\sum_{m=1}^L{\frac{1}{(\l -\xi - g \k_m)(\l +\xi-g \k_m)}}} \ .
\end{equation}
The limit $g\rightarrow 0$ reduces to the result in section~\ref{normalizationsection}. In an identical fashion, we can extend the analysis to the correlation functions obtained 
in section~\ref{evaluationcorrelationsection}. For instance, 
\begin{align}
& \granesperado{0}{\sigma^+_k \sigma^+_{k+1}}{\mu_1 \mu_2}_Z = \left[ \frac{\m_1+\xi-g\k_k}{\m_1-\xi-g \k_{k+1}} 
\frac{\m_2+\xi-g \k_{k+1}}{\m_2-\xi-g \k_k}-(\m_2 \leftrightarrow \m_1 ) \right] \nonumber \\ 
& \times \frac {1}{\big[ g(\k_{k+1}-\k_k)(\m_1 -\m_2 -i) \big]}\prod_{j=1}^{k+1}{\frac{\m_1 -\xi -g \k_j}{\m_1 +\xi -g \k_j} 
\frac{\m_2 -\xi -g \k_j}{\m_2 +\xi - g \k_j}} \ .
\end{align}
The norm is now given by
\begin{align}
	\sqrt{\braket{\mu_1 \mu_2}{\mu_1 \mu_2}_Z} & = \frac{2}{(\m_2 -\m_1)^2 -4\xi^2} \sum_j {\left[ \frac{1}{(\m_1-g \k_j)^2-4\xi^2} +\frac{1}{(\m_2-g \k_j)^2-4\xi^2} \right]} \nonumber \\
	& - \sum_j{\sum_k{\frac{1}{\big[ (\m_1-g \k_j)^2-4\xi^2 \big] \big[ ( \m_2-g \k_k )^2-4\xi^2 \big]}}} \ . 
\end{align}

We should stress that, in contrast with the homogeneous Heisenberg spin chain studied in the previous sections, the commutation of factors $(A+D)$ does not lead now to any indeterminate form because all the inhomogeneities involved are different. Therefore, we do not have 
to make use of the procedure we have  described in previous sections. For instance, the correlation function 
$\granesperado{0}{\sigma^+_k \sigma^+_{k+2}}{\mu_1 \mu_2}$ can be calculated by direct use of the commutation relations (\ref{commCAD}),
\begin{align}
& \granesperado{0}{\sigma^+_k \sigma^+_{k+2}}{\mu_1 \mu_2} = \granesperado{0}{C(\xi_k) (A+D) (\xi_{k+1}) C(\xi_{k+2}) B(\m_1) B(\m_2)}{0} p(k+2) \notag \\
& = \left[ \frac{\xi_{k}-\xi_{k+1}+i}{\xi_{k}-\xi_{k+1}} {\cal G}^{L}_0(k,k+2) + \frac{i}{\xi_{k+1}-\xi_k} {\cal G}^{L}_0(k+1,k+2) \right] p(k+2) \ ,
\label{k2}
\end{align}
where the correlation function ${\cal G}^{L}_{0}(k,l)=\langle 0 | C(\xi_k) C(\xi_l) | \mu_1 \mu_2 \rangle$ can be computed using 
expressions~(\ref{scalarproduct}) and (\ref{howtoscalarproduct}) for the scalar product. The factor $p(k+2)$, given by
\be
p(l)=\prod_{j=1}^{l}{\frac{\m_1 -\xi -g \k_j}{\m_1 +\xi -g \k_j} \frac{\m_2 -\xi -g \k_j}{\m_2 +\xi -g \k_j}} \ ,
\ee
collects the contribution from the momenta. \footnote{Note that if we take $g\rightarrow 0$, all inhomogeneities become the same, 
and (\ref{k2}) reproduces the limit in equation~(\ref{Fn1}).} 
We can in fact extend rather easily expression (\ref{k2}) to the case where the spin operators are located at arbitrary sites, 
$\langle 0 | \sigma^+_k \sigma^+_{l} | \mu_1 \mu_2 \rangle$. 
As all factors $(A+D)$ have different arguments, they can be trivially commuted. Therefore,
the correlation function must be invariant under exchange of the inhomogeneities, 
except for the factors coming from the correlators ${\cal G}^{L}_0(k,l)$. We find
\begin{align}
&\granesperado{0}{\sigma^+_k \sigma^+_{l}}{\mu_1 \mu_2} = \langle 0 \, |C(\xi_k) 
\prod_{j=k+1}^{l-1}{(A+D) (\xi_{j})} C(\xi_{l}) B(\m_1) B(\m_2) \, |0 \rangle \, p(l) = \notag\\
&= \left[ \, \prod_{j=k+1}^{l-1}{\frac{\xi_{k}-\xi_{j}+i}{\xi_{k}-\xi_{j}}} {\cal G}^{L}_0(k,l) 
+ \frac{-i}{\xi_{k}-\xi_{k+1}} \prod_{j=k+2}^{l-1}{\frac{\xi_{k+1}-\xi_{j}+i}{\xi_{k+1}-\xi_{j}}} {\cal G}^{L}_0(k+1,l) 
\right. \notag \\
& \left. + \left( \frac{\xi_{k+2}-\xi_{k+1}+i}{\xi_{k+2}-\xi_{k+1}} \right)
\frac{-i}{\xi_{k}-\xi_{k+2}} \prod_{j=k+3}^{l-1}{\frac{\xi_{k+2}-\xi_{j}+i}{\xi_{k+2}-\xi_{j}}} {\cal G}^{L}_0(k+2,l) + \dots \right] p(l) \ ,
\end{align}
or, after using the recursion relations,
\begin{align}
&\granesperado{0}{\sigma^+_k \sigma^+_{l}}{\mu_1 \mu_2}
= \left[ \, \prod_{m=k+1}^{l-1}{\frac{\xi_{k}-\xi_{m}+i}{\xi_{k}-\xi_{m}}} {\cal G}^{L}_0(k,l)  \: + \right. \notag \\
& \left. +\sum_{m=k+1}^{l-1}{ \left( \prod_{n=k+1}^{m-1}{\frac{\xi_{k}-\xi_{n}+i}{\xi_{k}-\xi_{n}}} \right) \frac{-i}{\xi_{k}-\xi_{m}} 
\left( \prod_{n=k+1}^{l-1}{\frac{\xi_{m}-\xi_{n}+i}{\xi_{m}-\xi_{n}}} \right) {\cal G}^{L}_0(m,l)} \right] p(l) \ .
\end{align}

A similar discussion holds in the case of higher order correlation functions, involving a larger amount of magnons. 
The analysis of the inhomogeneous case is in fact much less entangled than that of the homogeneous Heisenberg chain. We will, however, not present the resulting expressions in here.

%%%%%%%%%%%%%%%%%%%%%%%%%%%%%%%%%%%%%%%%%%%%%%%%%%%%%%%%%%%%%%%%%%
%%%%%%%%%%%%%%%%%%%%%%%%%%%%%%%%%%%%%%%%%%%%%%%%%%%%%%%%%%%%%%%%%%

\section{Concluding remarks} 
\label{conclusions}

In this article, we discuss the complexity of computing form factors and correlation functions of local spin operators both in the coordinate and the algebraic Bethe ansatz. 
We have focused our analysis on the $SU(2)$ Heisenberg spin chain, mostly due to its relevance for the $SU(2)$ sector of ${\cal N}=4$ supersymmetric Yang-Mills at one-loop. 
Nevertheless, we also dedicate an appendix to the other rank one sectors, i.e., $SL(2)$ and $SU(1,1)$.

We show that the main drawback of computing correlation functions using the CBA is the difficulty of the integrals and combinatorial sums that appear in the intermediate steps, 
which are not trivial to perform. For the ABA, computing correlation functions reduces to computing scalar products, which can be written in terms of a quotient of determinants 
in most of the cases, once we use the RTT relations to simplify them. However, this is not a straightforward step, only trivial when we have only one operator evaluated between Bethe states. 
Otherwise, we have to commute elements of the monodromy matrix with the same argument, which lead to expressions that contain a $\frac{0}{0}$ indeterminate form. We have discussed here how to solve them in some simple cases. In addition, we have dedicated the last section to the inhomogeneous Heisenberg spin chain, 
which can be applied to study ${\cal N}=4$ supersymmetric Yang-Mills beyond one-loop, and where the inhomogeneities prevent the indeterminate form we have mentioned.

Along our computations, and in order to compare results expected from the CBA with results in the ABA, special care was needed with 
the normalization of states. We have shown that agreement with results coming from the CBA requires that excitations in the spin chain must 
be defined as if we were using Zamolodchikov-Faddeev operators at $L\rightarrow \infty$. An important consequence of this is that states in the ABA pick up an S-matrix factor under the exchange 
of two rapidities. This property is crucial if we want, for instance, Watson's relations~\cite{Watson} to be satisfied by the form factors of the theory. 
An interesting continuation of our work in this article would be to understand what other constrains are imposed by the remaining axioms in Smirnov's 
form factor program~\cite{Smirnov}. The extension of Smirnov's program for relativistic integrable theories to worldsheet form factors for 
$AdS_5 \times S^5$ strings was discussed in~\cite{KM1,KM2}. In particular, it would be very interesting to understand the behaviour under crossing transformations of form factors 
evaluated using algebraic Bethe ansatz techniques. The crossing transformation corresponds to a shift in the rapidity by half the imaginary period 
of the torus that uniformises the magnon dispersion relation in the AdS/CFT correspondence~\cite{Janik}. However, one of the periods of the rapidity 
torus becomes infinitely large at weak coupling and, thus, both periodicity and crossing become invisible. It would be interesting if they can be recovered in the inhomogeneous Heisenberg spin chain. 
Nevertheless, most likely the 
dressing phase factor needs to be included, which we have not considered.

%%%%%%%%%%%%%%%%%%%%%%%%%%%%%%%%%%%%%%%%%%%%%%%%%%%%%%%%%%%%%%%%%%
%%%%%%%%%%%%%%%%%%%%%%%%%%%%%%%%%%%%%%%%%%%%%%%%%%%%%%%%%%%%%%%%%%

\vspace{8mm}

\centerline{\bf Acknowledgments}

\vspace{2mm}

\no
We are grateful to R. Ruiz and N. A. Slavnov for numerous discussions	and comments. The work of J.~M.~N. is supported by the Deutsche Forschungsgemeinschaft (DFG, German Research Foundation) under Germany’s Excellence Strategy – EXC 2121 “Quantum Universe” – 390833306.

\vspace{2mm}

%%%%%%%%%%%%%%%%%%%%%%%%%%%%%%%%%%%%%%%%%%%%%%%%%%%%%%%%%%%%%%%%%%
%%%%%%%%%%%%%%%%%%%%%%%%%%%%%%%%%%%%%%%%%%%%%%%%%%%%%%%%%%%%%%%%%%

\appendix

\renewcommand{\theequation}{\thesection.\arabic{equation}}
\csname @addtoreset\endcsname{equation}{section}

\section{General form of ${\cal F}^L_n$}
\label{A}
  
In this appendix we are going to obtain the general expression of the function ${\cal F}^L_n$. All along the calculation, the limit $\a \rightarrow 0$ will be assumed.
Using the first recurrence relation in (\ref{recurrenceeq}) and setting both $d$ and $\dpartial{d}{\lambda}$ to zero we find
\begin{equation}
	{\cal F}^L_n={\cal F}^L_0 + i\mathcal{D} {\cal F}^L_0 + i\mathcal{D} {\cal F}^L_1 + \dots + i\mathcal{D} {\cal F}^L_{n-1} \ .
\end{equation}
If we assume that $n<L-1$, the second recurrence equation gives 
\begin{equation}
\mathcal{D} {\cal F}^L_n=\binom{n}{0} \mathcal{D} {\cal F}^L_0+\binom{n}{1} \frac{i\mathcal{D}^2}{2!} 
{\cal F}^L_0+\dots=\sum_{j=0}^n{\binom{n}{j} \frac{i^j \mathcal{D}^{j+1}}{(j+1)!} {\cal F}^L_0} \ .
\end{equation}
Therefore we need to sum the series
\be
\sum_{j=0}^{n-1}{i\mathcal{D} {\cal F}^L_j}=\sum_{j=0}^{n-1}{\sum_{k=0}^j{\binom{j}{k} \frac{i^{k+1} \mathcal{D}^{k+1}}{(k+1)!} {\cal F}^L_0}} \ .
\ee
As a first step, we can commute the two sums as $\sum_{j=0}^{n-1}{\sum_{k=0}^j{}}=\sum_{k=1}^{n-1}{\sum_{j=k}^{n-1}{}}+\sum_{j=0}^{n-1}{\delta_{k,0}}$, 
because the index $j$ only appears in the limit of the sum and in the binomial coefficient, so it is easier to perform first the sum over $j$. 
The second term is easy to evaluate because we only have to calculate $\sum_{j=0}^{n-1}{\binom{j}{0}}=\binom{n}{1}$. 
The sum over $j$ of the first term can be computed using the properties of the binomial 
coefficients $\sum_{j=k}^{n-1}{\binom{j}{k}} = \binom{n-1+1}{k+1}$. 
The whole sum can thus be rewritten as
\begin{equation}
	{\cal F}^L_n={\cal F}^L_0+\sum_{k=1}^{n}{\binom{n}{k} \frac{i^{k} \mathcal{D}^{k}}{k!} {\cal F}^L_0} 
	= \sum_{k=0}^{n}{\binom{n}{k} \frac{i^{k} \mathcal{D}^{k}}{k!} {\cal F}^L_0} \ .
\end{equation}
This equation is true as long as $n < L-1$. If we want to get an expression valid for $n\geq L-1$, we have 
to take into account derivatives of $d$ of order greater than or equal to $L$, which can be performed 
independently of the calculation we have already done, because
\begin{displaymath}
	\mathcal{D}^{L+\alpha-1} {\cal F}^L_{j+1}=\frac{i}{L+\alpha} {\cal F}^L_j \dpartial[L+\alpha]{d}{\lambda} + \cdots \ ,
\end{displaymath}
where the dots stand for the part that we have already taken into account. Therefore,
the $d$-contribution to $\mathcal{D}{\cal F}^L_M$ will be of the form
\begin{align*}
	i\mathcal{D}{\cal F}^L_M &=\sum_{j=1}^{M-L+2}{\sum_{k=0}^{M+2-L-j}{\frac{i^{L+k-1}}{(L+k-1)!} 
	\binom{M-j}{L+k-2} \mathcal{D}^{L+k-1} {\cal F}^L_j}} \\
	&=\sum_{j=1}^{M-L+2}{\sum_{k=0}^{M+2-L-j}{\frac{i^{L+k}}{(L+k)!} \binom{M-j}{L+k-2} \dpartial[L+k]{d}{\lambda} {\cal F}^L_{j-1}}} \ ,
\end{align*}
and the derivative of $d$ can be calculated using Leibniz's rule,
\begin{displaymath}
	\left. \dpartial[L+k]{d}{\lambda} \right|_\xi =\sum_{j=0}^{L+k}{\binom{L+k}{j} 
	\dpartial[j]{(\lambda-\xi)^L}{\lambda} \, \dpartial[L+k-j]{(\lambda+\xi)^{-L}}{\lambda}} \ .
\end{displaymath}
Because we want to evaluate this expression at $\lambda=\xi$, the only non-zero contribution comes from the term 
with $L$ derivatives acting on $(\lambda - \xi)^L$. This implies
\begin{displaymath}
	\left. \dpartial[L+k]{d}{\lambda} \right|_\xi =\binom{L+k}{L} \dpartial[L]{(\lambda-\xi)^L}{\lambda} \, 
	\dpartial[k]{(\lambda+\xi)^{-L}}{\lambda}=\frac{(L+k)!}{k!} \, \frac{(L+k-1)!}{(L-1)!} \, \frac{(-1)^k}{i^{L+k}} \ .
\end{displaymath}
If we substitute that expression, we obtain
\begin{displaymath}
i\mathcal{D}{\cal F}^L_M =\sum_{j=1}^{M-L+2}{\sum_{k=0}^{M+2-L-j}{\binom{M-j}{L+k-2} \frac{(-1)^k (L+k-1)!}{(L-1)! k!} {\cal F}^L_{j-1}}} \ .
\end{displaymath}
After some algebra, the sum over $k$ takes the form
\begin{align*}
	&\sum_{k=0}^{m+2}{(-1)^k \binom{L+m}{L+k-2} \binom{L+k-1}{k}}=\frac{(L+m)!}{(L-1)!} 
	\sum_{k=0}^{m+2}{\frac{(L+k-1)!}{(L+k-2)!} \frac{(-1)^k}{(m-k+2)! k!}} \\
	&=\frac{(L+m)!}{(L-1)! (m+2)!} \sum_{k=0}^{m+2}{\left[ (-1)^k (L-1) \binom{m+2}{k}+(-1)^k k\binom{m+2}{k} \right]} \ ,
\end{align*}
where $m=M-L-j$. Properties of the binomial coefficients say that the first sum 
is zero (unless there is a single term, that is, if $m+2=0$) and the second sum is also zero 
(except if there are two terms, so that $m+2=1$). Then, the total contribution of these terms will amount to
\begin{align}
	& \sum_{M=L-1}^{n-1}{i\mathcal{D}{\cal F}^L_M}=\sum_{M=L-1}^{n-1}{\sum_{j=1}^{M-L+2}}{\frac{(M-j)!}{(L-1)! (M-L-j+2)!}} \nonumber \\
	& \times \big[ (L-1)\delta_{M-L-j+2,0} - (M-L-j+2) \delta_{M-L-j+2,1}) \big] {\cal F}^L_{j-1} \ , \end{align}
which telescopes, so that
\be
\sum_{M=L-1}^{n-1}{i\mathcal{D}{\cal F}^L_M} = {\cal F}^L_{n-L} \ . 
\ee
Therefore, the most general form of the correlation function ${\cal F}^L_n$ is 
\begin{equation}
{\cal F}^L_n=\sum_{k=0}^{n}{\binom{n}{k} \frac{i^{k} \mathcal{D}^{k}}{k!} {\cal F}^L_0}+\theta(n-L) {\cal F}^L_{n-L} \ . 
\label{FLn}
\end{equation}
where $\theta (x)$ is the Heaviside step function, with $\theta (x)=0$ if $x<0$ and $\theta (x)=1$ if $x \geq 0$. 

%%%%%%%%%%%%%%%%%%%%%%%%%%%%%%%%%%%%%%%%%%%%%%%%%%%%%%%%%%%%%%%%%%
%%%%%%%%%%%%%%%%%%%%%%%%%%%%%%%%%%%%%%%%%%%%%%%%%%%%%%%%%%%%%%%%%%

\section{Extension to $SL(2)$ and $SU(1|1)$ sectors}
\label{B}

Along most of this article, we have considered the $SU(2)$ spin $1/2$ homogeneous Heisenberg spin chain. 
The whole analysis that we have presented relies on the commutation relations of the elements of the monodromy matrix, 
equations~(\ref{commBAD}), (\ref{commCAD}) and (\ref{commCB}), the explicit form of the eigenvalues $a(\l)$ and $d(\l)$, and the S-matrix (and thus the Bethe equations). 
In this appendix, we want to give a taste of how the computations change for spin chains with symmetries $SL(2)$ and $SU(1|1)$.

\subsection{$SL(2)$ sector}

The case of the $SL(2)$ spin chain seems at first sight rather similar to the $SU(2)$ chain, 
because the commutation relationships between $(A+D)$ and $B$ are the same in both cases. 
However, the eigenvalues $a$ and $d$ are exchanged, 
\be
a_{SL(2)}(\l)=d_{SU(2)}(\l) \ , \quad d_{SL(2)}(\l)=a_{SU(2)}(\l)=1 \ . 
\ee
Fortunately, this does not prevent us from repeating the analysis that we have presented in section~\ref{evaluationcorrelationsection} to obtain 
for instance a set of recurrence equations for correlation 
functions with two spin operators, like equations~(\ref{recurrenceeq}) and~(\ref{GLn}). The derivation is just the same as in that section, but 
keeping terms in $a$ rather than terms in $d$. The final result is
\begin{align}
	{\cal F}^{L,(-1)}_{n+1}(\alpha)&= \big[1+a(\xi+\alpha)-i\left.\dpartial{a}{\lambda} \right|_{\xi+\alpha} \big] {\cal F}^{L,(-1)}_n (\alpha) 
	+ i\big[ a(\xi+\alpha)-1 \big] \mathcal{D} {\cal F}^{L,(-1)}_n (\alpha) \ , \nonumber \\
	\mathcal{D}^m {\cal F}^{L,(-1)}_{n+1} (\alpha)&=\left[ 1+a(\xi+\alpha) \right] 
	\mathcal{D}^m {\cal F}^{L,(-1)}_{n} (\alpha) - \frac{i}{m+1} \dpartial[m+1]{a}{\alpha} {\cal F}^{L,(-1)}_{n} (\alpha) \nonumber \\
	&+\frac{i}{m+1} \big[ a(\xi+\alpha) -1 \big] \mathcal{D}^{m+1} {\cal F}^{L,(-1)}_{n} (\alpha) \ , \\
	\mathcal{D}^m {\cal F}^{L,(-1)}_{0} (\alpha) &=\dpartial[m]{{\cal F}^{L,(-1)}_{0} (\alpha)}{\alpha} \ , \nonumber 
\end{align}
where the $(-1)$ superindex reminds that now we are calculating the correlation function in a $SL(2)$ spin chain. Finally, as in the $SU(2)$ sector,
\begin{equation}
	{\cal F}^{L,(-1)}_n =\sum_{k=0}^{n}{\binom{n}{k} \frac{(-i)^{k} \mathcal{D}^{k}}{k!} {\cal F}^{L,(-1)}_0}+\theta(n-L) {\cal F}^{L,(-1)}_{n-L} \ . 
\end{equation}

\subsection{$SU(1|1)$ sector}

The case of a $SU(1|1)$ spin chain is slightly more complex to handle because of the grading of the algebra. 
However, the evaluation of correlation functions turns out to be simpler than in the $SU(2)$ and $SL(2)$ sectors. The commutation relations are given by~\cite{BR},
\begin{align}
	B(\m) B(\l) &=-\frac{\m-\l+i}{\m-\l-i} B(\l) B(\m) \ , 	\nonumber \\
	A (\mu) B(\lambda) &=\left( 1+\frac{i}{\m-\l} \right) B(\lambda) A(\mu) +\frac{i}{\m-\l} B(\mu) A (\lambda)  \ , 	\nonumber \\
	D (\mu) B(\lambda) &=\left( 1+\frac{i}{\m-\l} \right) B(\lambda) D(\mu) +\frac{i}{\m-\l} B(\mu) D (\lambda)  \ , 	\nonumber \\
	C(\lambda) A (\mu) &=\left( 1+\frac{i}{\l-\m} \right) A(\mu) C(\lambda) -\frac{i}{\l -\m} A (\lambda) C(\mu) \ , 	\nonumber \\
	C(\lambda) D (\mu) &=\left( 1+\frac{i}{\l-\m} \right)  D(\mu) C(\lambda) -\frac{i}{\l -\m} D (\lambda) C(\mu) \ .
\end{align}
These commutation relations present some differences with respect to their $SU(2)$ counterparts. 
The most important one is that they have the same form both for $A$ and~$D$. 
Another important difference is that the transfer matrix has to be graded and thus new we have $T(\l)=A(\l)-D(\l)$ 
instead of $A(\l)+D(\l)$. On the contrary, the form of the functions $a$ and $d$ does not change.

We can now follow section~\ref{evaluationcorrelationsection} and 
proceed to find the commutation relations between the transfer matrix and the $C$ operators in the limit where the rapidities are equal. We obtain
\begin{align}
	&\lim_{\b \rightarrow \a} C(\a) (A-D)(\b)=\lim_{\b \rightarrow \a} \frac{\a-\b +i}{\a-\b}  (A-D)(\b) C(\a)-\frac{i}{\a-\b} (A-D)(\a) C(\b) \notag \\
	&=(A-D)(\a) C(\a) +i \left[ (A-D)(\a) \left.\dpartial{C(\l)}{\l} \right|_{\l=\a} - \left.\dpartial{(A-D)(\l)}{\l} \right|_{\l=\a} C(\a) \right] \ .
\end{align}
We can also calculate the derivatives and thus the recurrence relations become
\begin{align}
	{\cal F}^{L,(0)}_{n+1} (\a) &=(1-d+i\partial d) {\cal F}^{L,(0)}_n (\a) +i (1-d) {\cal D} {\cal F}^{L,(0)}_n (\a) \notag \ , \\
	{\cal D}^m {\cal F}^{L,(0)}_{n+1} (\a) &= (1-d) {\cal D}^m {\cal F}^{L,(0)}_n (\a) \notag \\
	& +\frac{i}{m+1} \left[ (1-d) {\cal D}^{m+1} {\cal F}^{L,(0)}_n (\a) +\dpartial[m+1]{d}{\a} {\cal F}^{L,(0)}_n (\a) \right] \notag \ , \\
	{\cal D}^m {\cal F}^{L,(0)}_0 (\a) &=\dpartial[m]{{\cal F}^{L,(0)}_0 (\a)}{\a} \ ,
\end{align}
where the $(0)$ superindex states that now we are calculating a correlation function in the case of an $SU(1|1)$ spin chain, and where, as usual 
$d=d(\xi+\alpha)$ and $\partial d=\left. \dpartial{d(\l)}{\l} \right|_{\l=\xi+\a}$. These recurrence equations can again be written in terms 
of some starting condition corresponding to $n=0$, using
\begin{equation}
{\cal F}^{L,(0)}_n=\sum_{k=0}^{n}{\binom{n}{k} \frac{i^{k} \mathcal{D}^{k}}{k!} {\cal F}^{L,(0)}_0} \ ,
\end{equation}
provided we keep $n<L-1$.

%%%%%%%%%%%%%%%%%%%%%%%%%%%%%%%%%%%%%%%%%%%%%%%%%%%%%%%%%%%%%%%%%%
%%%%%%%%%%%%%%%%%%%%%%%%%%%%%%%%%%%%%%%%%%%%%%%%%%%%%%%%%%%%%%%%%%

\section{Correlation functions involving states with three magnons}
\label{C}

Computing correlation functions using the ABA becomes a challenge when the number of magnons increases. In this appendix, we give a taste of how the complexity 
increases for the case of correlations functions involving states with three magnons. 

There are four non-vanishing correlation functions whose most populated state contains three magnons.
The first one is just the scalar product $\braket{\l _1 ,\l _2 ,\l _3}{\m_1 ,\m_2 ,\m_3}$ and can be directly calculated using the 
Gaudin formula~(\ref{gaudin}). The second one is the form factor of a single spin operator,
$\granesperado{\l _1 ,\l _2 }{\sigma^+_k}{\m_1 ,\m_2 ,\m_3}$, which can be evaluated 
in a straightforward extension of the computation of $\granesperado{\lambda}{\sigma^+_k}{\mu_1 \mu_2}$ in section~\ref{lambdasigmamumusection}. 
The third kind of correlation function involving three magnons is 
$\granesperado{\lambda}{\sigma^+_k \sigma^+_l}{\m _1 \m _2 \m _3}$, and the fourth one is $\granesperado{0}{\sigma^+_k \sigma^+_l \sigma^+_m}{\m _1 \m _2 \m _3}$.
These last two types of correlators are the ones that we will consider in this appendix. We will start studying the third one (the one that involves two spin operators), and along the computation we will find that it 
involves correlation functions of the form $\granesperado{0}{\sigma^+_k \sigma^+_{k+1}  \sigma^+_{k+n+2}}{\m _1 \m _2 \m _3}$, which are a particular case of the fourth 
type of correlator. 

We will start by writing the problem in terms of operators of the monodromy matrix using (\ref{sigmaplus}),
\begin{multline}
	\langle \lambda |\sigma^+_k \sigma^+_l |\mu_1 \mu_2 \mu_3\rangle =\\
	 \granesperado{0}{C(\l )(A+D)^{k-1} (\xi) C(\xi) (A+D)^{n} (\xi) C(\xi) (A+D)^{L-l} (\xi) }{\mu_1 \mu_2 \mu_3} \ ,
\end{multline}
where, as before, $n=L+l-k-1$. 
The factor $(A+D)^{k-1}$ acts on $C(\l )$ to give $e^{-ip_\l (k-1)}$, and the factor $(A+D)^{L-l}$ 
acts on the three-magnon state to give $e^{-i(p_1+p_2+p_3)\cdot (L-l)}=e^{i(p_1+p_2+p_3)l}$, 
where we have used the periodicity condition for the Bethe roots.
Therefore, our main problem will be to find the correlation function
\be 
{\cal H}^L_{n}(\a)=\granesperado{0}{C(\l ) C(\xi+\alpha) (A+D)^n (\xi) C(\xi) B(\m _1) B(\m _2) B(\m _3)}{0} \ .
\ee
Following the procedure that we have constructed along the article, this can be done by relating ${\cal H}^L_{n+1}(\a)$ to ${\cal H}^L_{n}(\a)$. In order to do this, 
let us start by introducing 
\be
{\cal H}^L_{n+1} (\l ,\a, \delta)=\lim_{\b\rightarrow \a} \granesperado{0}{C(\l ) C(\xi +\a) (A+D)(\xi+\beta) {\cal O} (\d)}{0} \ .
\ee
Now we just need to apply the commutation relations~(\ref{commCAD}) two times in each step, which gives
\begin{align}
	&{\cal H}^L_{n+1} (\l ,\a, \delta) = \lim_{\b \rightarrow \a} \Big{\{}\big[ 1+d(\xi+\b) \big] {\cal H}^L_{n} (\l ,\a, \delta) \nonumber \\
	&-\frac{i}{\l -\xi -\b} \left[ (d(\xi+\b)-1) {\cal H}^L_{n} (\l ,\a, \delta)-(d(\l)-1) {\cal H}^L_{n} (\xi +\b ,\a, \delta) \right] \nonumber \\
	&-\frac{i}{\a -\b} \left[ (d(\xi+\b) -1) {\cal H}^L_{n} (\l ,\a, \delta) -(d(\xi+\a)-1) {\cal H}^L_{n} (\l ,\b, \delta) \right] \nonumber \\
	&+\frac{i}{\a -\b} \ \frac{i}{\l -\xi -\b} \left[ (d( \xi+\b)+1) {\cal H}^L_{n} (\l ,\a, \delta) -(d(\l )+1) {\cal H}^L_{n} (\xi+\b ,\a, \delta) \right] \nonumber \\
	&-\frac{i}{\a -\b} \ \frac{i}{\l -\xi -\a} \left[ (d( \xi+\a)+1) {\cal H}^L_{n} (\l ,\b, \delta) -(d(\l )+1) {\cal H}^L_{n} (\xi+\b ,\a, \delta) \right] \Big{\}} \ . \label{Hpoles}
\end{align}
Taking the limit and applying the Bethe equation for the rapidity $\l$ we obtain
\begin{align}
	&{\cal H}^L_{n+1} (\l ,\a, \delta)=\Big( 1+d+i\partial d+\frac{\partial d-i(d-1)}{\l -\xi -\a}+\frac{d+1}{(\l-\xi-\a)^2} \Big) {\cal H}^L_{n} (\l ,\a, \delta) \notag \\
	&+\Big[ i(1-d)-\frac{d+1}{\l -\xi -\a} \Big] \dpartial{{\cal H}^L_{n} (\l ,\a, \delta)}{\a} -\frac{2}{(\l -\xi -\a)^2} {\cal H}^L_{n} (\xi+\a ,\a, \delta) \ ,
\label{Hn+1}
\end{align}
where, as before, $d=d(\xi+\alpha)$ and $\partial d=\left. \dpartial{d}{\lambda} \right|_{\xi +\alpha}$. 
The next step of the calculation is a little bit more involved than in the previous cases because, according to (\ref{Hn+1}), information about  
both functions ${\cal H}^L_{n} (\l ,\a, \delta)$ and ${\cal H}^L_{n+1} (\xi+\a ,\a, \delta)$ is now needed. 
This will make the computation slightly more difficult, but still manageable. In the expressions 
below, we will define ${\cal H}^L_{n+1} (\a+\xi ,\a, \delta)=\hat{{\cal H}}^L_{n+1} (\a ,\a, \delta)$ for convenience. This function $\hat{\cal H}$ has a nice interpretation because
\begin{align}
	\granesperado{0}{\sigma^+_k \sigma^+_{k+1}  \sigma^+_{k+n+2}}{\m _1 \m _2 \m _3} 
	&=\granesperado{0}{C(\xi ) C(\xi) (A+D)^{n} (\xi) C(\xi) (A+D)^{L-n-k-2} (\xi) }{\mu_1 \mu_2 \mu_3} \notag \\
	&=\hat{\cal H}^L_{n} e^{i(p_1+p_2+p_3)(n+k+2)} \ .
\end{align}
Our starting point is thus to find the recursive equation for $\hat{\cal H}$. This can be obtained setting $\l=\g+\xi$ in expression (\ref{Hpoles}) 
and taking the limit $\g \rightarrow \a$,
\begin{align}
&\hat{\cal H}^L_{n+1} (\a ,\a, \delta)=\lim _{\b\rightarrow\a}\frac{1}{\b-\a} \Big[ (1+d) 
\left. \dpartial{\hat{\cal H}^L_{n} (\l ,\a, \delta)}{\l} \right|_{\l=\a} \!\!
- (1+d) \left. \dpartial{\hat{\cal H}^L_{n} (\a ,\l, \delta)}{\l} \right|_{\l=\a} \Big] \notag \\
& + \big(1+d+2i\partial d-\frac{1}{2} \partial^2 d \big) \, \hat{\cal H}^L_{n} (\a ,\a, \delta) + \big[ 2i(1-d) +\partial d \big] 
\left. \dpartial{\hat{\cal H}^L_{n} (\l ,\a, \delta)}{\l} \right|_{\l=\a}  \notag \\
& + \frac{1+d}{2} \left. \dpartial[2]{\hat{\cal H}^L_{n} (\l ,\a, \delta)}{\l} \right|_{\l=\a}-(1+d) 
\left. \frac{\partial^2 \hat{\cal H}^L_{n} (\l _1 ,\l _2, \delta)}{\partial \l_1 \partial \l_2} \right|_{\substack{\l_{1} = \a \\ \l_{2} = \a}} \ . 
\label{c7}
\end{align}
Note that, although the first term in this expression seems divergent, it vanishes because of the commutator of two $C$ operators vanishes, 
which makes the two derivatives equal. 
However, this way of calculating recursively $\hat{\cal H} (\a , \a , \d)$ is going to create more problem than it solves, 
as it will imply calculating the recurrence equation of the derivative of $\hat{\cal H} (\l ,\m)$ with respect to either the first or the second argument. 
Therefore, we are going to present the recursion relation of $\hat{\cal H} (\b , \a , \d)$ but without taking the limit $\b\rightarrow \a$. 
To obtain this recurrence relation, we only need to substitute $\l=\xi +\b$ in equation~(\ref{Hn+1}), but {\em without} imposing $d(\l)=1$,
\begin{align}
& \hat{\cal H}^L_{n+1} (\b ,\a, \delta)=\Big( 1+d+i\partial d+\frac{\partial d-i(d-1)}{\b -\a}+\frac{d+1}{(\b-\a)^2} \Big) \hat{\cal H}^L_{n} (\b ,\a, \delta) \notag \\
& +\Big[ i(1-d)-\frac{d+1}{\b -\a} \Big] \dpartial{\hat{\cal H}^L_{n} (\b ,\a, \delta)}{\a} +\left[ \frac{i(d'-1)}{\b-\a} -\frac{d'+1}{(\b -\a)^2} \right] 
\lim_{\g \rightarrow \a} \hat{\cal H}^L_{n} (\g ,\a, \delta) \ ,
\label{hatHn+1}
\end{align}
where $d'=d(\xi +\b)$. Note that if we take $\b \rightarrow \a$ equation (\ref{hatHn+1}) gives (\ref{c7}). Now in the recurrence relation we need to include 
$\lim_{\g \rightarrow \a} \hat{\cal H}^L_{n} (\g ,\a, \delta)$, but this quantity is obviously known once we know $\hat{\cal H}^L_{n} (\b ,\a, \delta)$.

We also need a recurrence equation for the derivatives. For the case of ${\cal H}_{n}$ we have
\begin{align}
	&{\cal D}^n {\cal H}^L_{m+1} (\l ,\a, \delta)=(1+d) {\cal D}^n {\cal H}^L_{m} -\frac{i}{\l -\xi -\a} \left[ (d-1) {\cal D}^n {\cal H}^L_{m} \right] \notag \\
	&+\frac{i}{n+1} (\partial^{n+1} d) {\cal H}^L_{m}+(1-d)\frac{i}{n+1} {\cal D}^{n+1} {\cal H}^L_{m} \notag\\
	&+\sum_{k=0}^n \sum_{l=0}^{k+1}{\frac{n!}{(n-k)! (k+1-l)!} \frac{1}{(\l -\xi -\a)^{l+1}} \left[ \partial^{k+1-l} (d+1) {\cal D}^{n-k} {\cal H}^L_{m} \right.} \notag \\ 
	&\left. -(d(\l )+1) {\cal D}^{n-k}_1 {\cal D}^{k+1-l}_2 \hat{\cal H}^L_{m} \right] \notag \\
	&-\sum_{k=0}^n{\sum_{l=0}^{n-k}{ \frac{n!}{(k+1)! (n-l-k)!} \frac{1}{(\l -\xi -\a)^{l+1}} \left[ \partial^{n-k-l}(d+1) {\cal D}^{k+1} {\cal H}^L_{m} \right.}} \notag \\
	&\left. -(d(\l) +1) {\cal D}^{n-k-l}_1 {\cal D}^{k+1}_2 \hat{\cal H}^L_{m} \right] \ .
\end{align}
The last two sums cancel themselves except for the terms with $k=n$. Therefore
\begin{align}
	&{\cal D}^n {\cal H}^L_{m+1} (\l ,\a, \delta)=(1+d) {\cal D}^n {\cal H}^L_{m} -\frac{i}{\l -\xi -\a}  (d-1) {\cal D}^n {\cal H}^L_{m} \notag \\
	&+\frac{i}{n+1} (\partial^{n+1} d) {\cal H}^L_{m}+(1-d)\frac{i}{n+1} {\cal D}^{n+1} {\cal H}^L_{m} \notag\\
	&+\sum_{l=0}^{n+1}{\frac{n!}{(n+1-l)!} \frac{1}{(\l -\xi -\a)^{l+1}} \left[ \partial^{n+1-l} (d+1) {\cal H}^L_{m}  - 2 \, {\cal D}^{n+1-l} \hat{\cal H}^L_{m} \right]} \notag \\
	&-\frac{1}{n+1} \frac{1}{(\l -\xi -\a)} \left[ (d+1) {\cal D}^{n+1} {\cal H}^L_{m} - 2 \, {\cal D}^{n+1} \hat{\cal H}^L_{m} \right] \ ,
\end{align}
where we have used that $d(\l)=1$. In a similar way, we can obtain an expression for the derivatives of $\hat{\cal H}$,
\begin{align}
	&{\cal D}^n \hat{\cal H}^L_{m+1} (\b ,\a, \delta)=(1+d) {\cal D}^n \hat{\cal H}^L_{m} -\frac{i}{\b-\a}  (d-1) {\cal D}^n \hat{\cal H}^L_{m} \notag \\
	&+\frac{i}{n+1} (\partial^{n+1} d) \hat{\cal H}^L_{m}+(1-d)\frac{i}{n+1} {\cal D}^{n+1} \hat{\cal H}^L_{m} +\frac{i}{\b -\a} (d' -1) 
	\lim_{\g \rightarrow \a} \dpartial[n]{\hat{\cal H}^L_{n} (\g ,\a, \delta)}{\a} \notag\\
	&+\sum_{l=0}^{n+1}{\frac{n!}{(n+1-l)!} \frac{1}{(\b -\a)^{l+1}} \left[ \partial^{n+1-l} (d+1) \hat{\cal H}^L_{m}  - (d'+1)\, 
	\lim_{\g \rightarrow \a} \dpartial[{n+1-l}]{\hat{\cal H}^L_{m} (\g ,\a ,\d)}{\a} \right]} \notag \\
	&-\frac{1}{n+1} \frac{1}{(\b -\a)} \left[ (d+1) {\cal D}^{n+1} \hat{\cal H}^L_{m} - (d'+1) \, \lim_{\g \rightarrow \a} 
	\dpartial[{n+1}]{\hat{\cal H}^L_{m} (\g ,\a ,\d)}{\a} \right] \ .
\end{align}

At this point the problem is, at least formally, solved. We have found the recursion relation for $\hat{\cal H}$ and its derivatives, 
with $\granesperado{0}{C(\xi +\b) C(\xi+\a) (\xi) C(\xi)}{\mu_1 \mu_2 \mu_3} =\hat{\cal H}^L_{n} (\b , \a)$ as the initial condition. 
These functions can then be substituted in the recursion relation for $\cal H$ and thus we can obtain the desired correlation function. 
However, we are not going to present the general form for the correlation function ${\cal H}^L_n$ as a function of ${\cal H}^L_0$, $\hat{\cal H}^L_0$ 
and their derivatives, because it becomes rather lengthy. This is because when we substitute 
the expression for the derivatives, the recursion relations turn out to depend on all the ${\cal H}_i$ with $0\leq i\leq n$, even 
after we take the limit $\a \rightarrow 0$. Instead, we can present the case of correlation functions with $n$ small, 
to exhibit the nested procedure to write the result in terms of the initial functions ${\cal H}^L_0$ and $\hat{\cal H}^L_0$. 
In particular, we are going to consider the first three functions, with $n=1$, $n=2$ and $n=3$. Thus, we can safely assume that $n<L-1$ 
so that all the $d$ and $\partial^k d$ factors can be set to zero in the limit $\a \rightarrow 0$. The first of these correlation functions is given by
\be 
{\cal H}^L_{1} = \big( 1+ i c(\l) + c(\l)^2 \big) \, {\cal H}^L_0 + \big( i-c(\l) \big) \left. \dpartial{{\cal H}^L_0 (\l ,\a)}{\a} \right| _{\a =0} 
- 2 c(\l)^2 \, \hat{\cal H}^L_0 \ ,
\ee
where, for convenience, we have defined $c(\l) = 1/(\l-\xi)$. For simplicity, if no arguments of these functions are given, ${\cal H}^L (\l, 0) $ 
and $\hat{\cal H}^L  (0, 0)$ must be understood. The last step of the computation reduces to calculating some initial conditions, which now are
\begin{align}
	{\cal H}^L_0 (\l, \a) &=\granesperado{0}{C(\l ) C(\xi+\alpha) C(\xi) B(\m _1) B(\m _2) B(\m _3)}{0} \ , \\
	\hat{\cal H}^L_0  (\a, \b)&= {\cal H}^L_0 (\xi+\a, \b) \ .
\end{align}
These functions can be easily computed using equations~(\ref{howtoscalarproduct}). However, we are not going to present the explicit expression 
for these scalar products because of its length and because we want to show the way to solve the recurrence relation rather 
than obtaining the explicit value of the correlation function.

The functional dependence of ${\cal H}_1$ on ${\cal H}_0$ is repeated for a given value of $n$ and the lower correlator. That is, 
in the limit $\a \rightarrow 0$ the recurrence relation for ${\cal H}^L_{n+1}$ is given by
\be 
{\cal H}^L_{n+1} = \big( 1+ i c(\l) + c(\l)^2 \big) {\cal H}^L_n + \big( i-c(\l) \big) {\cal D} {\cal H}^L_n - 2 c(\l)^2 \hat{\cal H}_n \, \ .
\ee
Therefore, for the second correlation function, we have
\be
{\cal H}^L_{2} = \big( 1+ i c(\l) + c(\l)^2 \big) {\cal H}^L_1 + \big( i-c(\l) \big) {\cal D} {\cal H}^L_1 - 2 c(\l)^2 \hat{\cal H}_1 \ .
\label{c16}
\ee
As we already know ${\cal H}^L_1$, it only remains to find the other two functions entering (\ref{c16}). This can be done using the equations that we have obtained 
along this appendix. We get
\begin{align}
{\cal D} {\cal H}^L_{1}&=c(\l)^3 {\cal H}^L_0+(1+ic(\l))  \left. \dpartial{{\cal H}^L_0 (\l ,\a)}{\a} \right| _{\a =0}+\frac{i(1+ic(\l))}{2}  \left. \dpartial[2]{{\cal H}^L_0 (\l ,\a)}{\a} \right| _{\a =0} \notag \\
&-2c(\l)^3 \hat{\cal H}^L_0 -2c(\l)^2 \left. \dpartial{\hat{\cal H}^L_0 (0 ,\a)}{\a} \right| _{\a =0} \ , \\
\hat{\cal H}^L_1 (\b , 0) &=  \Big( 1+\frac{i}{\b}+\frac{1}{\b^2} \Big) \hat{\cal H}^L_0 (\b , 0) +\Big[ i-\frac{1}{\b } \Big] 
\left. \dpartial{\hat{\cal H}^L_0 (\b ,\a)}{\a} \right|_{\a=0}-\left[ \frac{i}{\b} +\frac{1}{\b^2} \right] \hat{\cal H}^L_{0} \ , \\
\hat{\cal H}^L_1 &=  \hat{\cal H}^L_{0} +2i  \left. \dpartial{\hat{\cal H}^L_0 (0 ,\a)}{\a} \right| _{\a =0} +\frac{1}{2} 
\left. \dpartial[2]{\hat{\cal H}^L_0 (0 ,\a)}{\a} \right| _{\a =0}-\left. \frac{ \partial ^2 \hat{\cal H}^L_0 (\a ,\b)}{\partial \a \partial \b} \right| _{\substack{\a =0 \\ \b=0}} \ ,
\end{align}
which reduce again to some dependence on the initial conditions we have described before.

An identical computation can be done for ${\cal H}^L_{3}$,
\be
{\cal H}^L_{3} = \big( 1+ i c(\l) + c(\l)^2 \big) {\cal H}^L_2 + \big( i-c(\l) \big) {\cal D} {\cal H}^L_2 - 2 c(\l)^2 \hat{\cal H}_2 \ .
\ee
Now, besides ${\cal H}^L_{2}$, that has been calculated just before, we need
\begin{align}
{\cal D} {\cal H}^L_{2}&=c(\l)^3 {\cal H}^L_1+(1+ic(\l))  {\cal D} {\cal H}^L_{1}+\frac{i(1+ic(\l))}{2}  {\cal D}^ 2 {\cal H}^L_{1} \notag \\
&-2c(\l)^2 \Big( c(\l) \hat{\cal H}^L_1 +{\cal D} \hat{\cal H}^L_{1} \Big) \ , \\
{\cal D}^2 {\cal H}^L_{1}&=2 c(\l)^4 {\cal H}^L_0+(1+ic(\l))  \left. \dpartial[2]{{\cal H}^L_0 (\l ,\a)}{\a} \right| _{\a =0}+\frac{i(1+ic(\l))}{3}  
\left. \dpartial[3]{{\cal H}^L_0 (\l ,\a)}{\a} \right| _{\a =0} \notag \\
&-4c(\l)^4  \hat{\cal H}^L_0 -4 c(\l)^3 \left. \dpartial{\hat{\cal H}^L_0 (0 ,\a)}{\a} \right| _{\a =0} -2c(\l)^2 \left. \dpartial[2]{\hat{\cal H}^L_0 (0 ,\a)}{\a} \right| _{\a =0} \ ,\\
{\cal D} \hat{\cal H}^L_{1} &=\left. \dpartial{\hat{\cal H}^L_0 (0 ,\a)}{\a} \right| _{\a =0} 
+ \frac{i}{2}\left. \dpartial[2]{\hat{\cal H}^L_0 (0 ,\a)}{\a} \right| _{\a =0}+i\left. \frac{ \partial ^2 \hat{\cal H}^L_0 (\a ,\b)}{\partial \a \partial \b} 
\right| _{\substack{\a =0 \\ \b=0}} \notag \\
& + \frac{1}{3!} \left. \dpartial[3]{\hat{\cal H}^L_0 (0 ,\a)}{\a} \right| _{\a =0} 
- \frac{1}{2} \left. \frac{ \partial ^3 \hat{\cal H}^L_0 (\a ,\b)}{\partial \a \partial \b^ 2} \right| _{\substack{\a =0 \\ \b=0}} \ , \\
\hat{\cal H}^L_{2} &=\hat{\cal H}^L_{0} +4i \left. \dpartial{\hat{\cal H}^L_0 (0 ,\a)}{\a} \right| _{\a =0} -4 
\left. \frac{ \partial ^2 \hat{\cal H}^L_0 (\a ,\b)}{\partial \a \partial \b} \right| _{\substack{\a =0 \\ \b=0}}+\frac{i}{2}\left. 
\dpartial[3]{\hat{\cal H}^L_0 (0 ,\a)}{\a} \right| _{\a =0} \notag \\
& - \frac{3i}{2} \left. \frac{ \partial ^3 \hat{\cal H}^L_0 (\a ,\b)}{\partial \a \partial \b^2} \right| _{\substack{\a =0 \\ \b=0}} 
-\frac{1}{3!} \left. \frac{ \partial ^4 \hat{\cal H}^L_0 (\a ,\b)}{\partial \a \partial \b^3} \right| _{\substack{\a =0 \\ \b=0}} +\frac{1}{2!^2} 
\left. \frac{ \partial ^4 \hat{\cal H}^L_0 (\a ,\b)}{\partial \a^2 \partial \b^2} \right| _{\substack{\a =0 \\ \b=0}} \ .
\end{align}

The cases with higher values of $n$ can be obtained along similar lines. 

To conclude our analysis, we will brief comment on the calculation of correlation functions $\granesperado{0}{\s^+_k \s^+_l \s^+_m}{\{ \m \}}$, 
with general values of $k$, $l $ and $m$. In this case, the value of $n$ in $\hat{\cal H}^L_n$ will be proportional to the separation of $l$ and $m$. 
But it still remains to separate $k$ from $l$. This last step can be solved using the tools from section~\ref{0sigmasigmamumu}, because the problem 
in both cases is the same.

%%%%%%%%%%%%%%%%%%%%%%%%%%%%%%%%%%%%%%%%%%%%%%%%%%%%%%%%%%%%%%%%%%
%%%%%%%%%%%%%%%%%%%%%%%%%%%%%%%%%%%%%%%%%%%%%%%%%%%%%%%%%%%%%%%%%%


\begin{thebibliography}{99}

\renewcommand{\baselinestretch}{.99}
\normalsize

\bibitem{Bethe} H.~Bethe,
{\em On the theory of metals. 1. Eigenvalues and eigenfunctions for the linear atomic chain},
Z. Phys. \textbf{71} (1931), 205-226.

\bibitem{ABA} L.~D.~Faddeev, E.~K.~Sklyanin and L.~A.~Takhtajan,
{\em The Quantum Inverse Problem Method. 1},
Teor. Mat. Fiz. \textbf{40} (1979), 194-220.

\bibitem{Baxter} R.~J.~Baxter,
{\em Partition function of the Eight-Vertex lattice model},
Annals of Physics \textbf{70} 1 (1972), 193-228.

\bibitem{Reshetikhin} N.~Yu.~Reshetikhin,
{\em The functional equation method in the theory of exactly soluble quantum system},
Sov. Phys. JETP \textbf{57} (1983), 691-696.

\bibitem{KMT} N.~Kitanine, J.~M.~Maillet and V.~Terras,
{\em Form factors of the XXZ Heisenberg spin-$\frac{1}{2}$ finite chain},
Nucl.~Phys.~B 554 (1999), \textbf{3}, 647 {\tt [arXiv:9807020 [math-ph]]}.
%%CITATION = ARXIV:980720;%%

\bibitem{Kitaninereview} N.~Kitanine, J.~M.~Maillet, N.~A.~Slavnov, and V.~Terras,
{\em On the algebraic Bethe ansatz approach to the correlation functions
of the XXZ spin-1/2 Heisenberg chain}, in {\em Solvable
Lattice Models 2004 - Recent Progress on Solvable Lattice Models},
RIMS, Kyoto, {\bf 1480} (2006) 14,  {\tt [arXiv:0505006 [hep-th]]}.
%%CITATION = HEP-TH/0505006;%%

\bibitem{Korepin} V.~E.~Korepin,
{\em Calculation of norms of Bethe wave functions},
Commun. Math. Phys. \textbf{86} (1982), 391.
%%CITATION = CMPHA,86,391;%%

\bibitem{Slavnov} N.~A.~Slavnov, 
{\em Calculation of scalar products of wave functions and form factors in the framework of the algebraic Bethe ansatz},
Theor. Math. Phys. {\bf 79} (1989) 502. 

\bibitem{Beisertreview} N.~Beisert, C.~Ahn, L.~F.~Alday, Z.~Bajnok, J.~M.~Drummond, L.~Freyhult, N.~Gromov and R.~A.~Janik {\it et al.},
{\em Review of AdS/CFT Integrability: An Overview},
Lett.\ Math.\ Phys.\  {\bf 99} (2012) 3, 
{\tt [arXiv:1012.3982 [hep-th]]}.
%%CITATION = ARXIV:1012.3982;%%

\bibitem{bootstrap1}
R.~Rattazzi, V.~S.~Rychkov, E.~Tonni and A.~Vichi,
{\em Bounding scalar operator dimensions in 4D CFT},
JHEP \textbf{12} (2008), 031.
{\tt [arXiv:0807.0004 [hep-th]]}.
%%CITATION = ARXIV:0807.0004;%%

\bibitem{bootstrap2}
S.~El-Showk, M.~F.~Paulos, D.~Poland, S.~Rychkov, D.~Simmons-Duffin and A.~Vichi,
{\em Solving the 3D Ising Model with the Conformal Bootstrap},
Phys. Rev. D \textbf{86} (2012), 025022,
{\tt [arXiv:1203.6064 [hep-th]]}.
%%CITATION = ARXIV:1203.6064;%%

\bibitem{hexagon1}
B.~Basso, S.~Komatsu and P.~Vieira,
{\em Structure Constants and Integrable Bootstrap in Planar N=4 SYM Theory},
{\tt [arXiv:1505.06745 [hep-th]]}.
%%CITATION = ARXIV:1505.0674;%%

\bibitem{hexagon2}
B.~Eden and A.~Sfondrini,
{\em Tessellating cushions: four-point functions in $\mathcal{N} $ = 4 SYM},
JHEP \textbf{10} (2017), 098, 
{\tt [arXiv:1611.05436 [hep-th]]}.
%%CITATION = ARXIV:1611.05436;%%

\bibitem{hexagon3}
T.~Fleury and S.~Komatsu,
{\em Hexagonalization of Correlation Functions},
JHEP \textbf{01} (2017), 130, 
{\tt [arXiv:1611.05577 [hep-th]]}.
%%CITATION = ARXIV:1611.05577;%%

\bibitem{RV} R.~Roiban and A.~Volovich,
{\em Yang-Mills correlation functions from integrable spin chains},
JHEP {\bf 0409} (2004) 032, {\tt [arXiv:0407140 [hep-th]]}.
%%CITATION = HEP-TH/0407140;%%

\bibitem{Escobedo} J.~Escobedo, N.~Gromov, A.~Sever and P.~Vieira,
{\em Tailoring Three-Point Functions and Integrability},
JHEP \textbf{1109} (2011), 028, {\tt [arXiv:1012.2475 [hep-th]]}.
%%CITATION = ARXIV:1012.2475;%%

\bibitem{Foda} O.~Foda,
{\em ${\cal N}=4$ SYM structure constants as determinants},
JHEP {\bf 1203} (2012) 096, 
{\tt [arXiv:1111.4663 [math-ph]]}.
%%CITATION = ARXIV:1111.4663;%%

\bibitem{Kostov} I.~Kostov,
{\em Classical limit of the three-point function of ${\cal N}=4$ supersymmetric Yang-Mills theory from integrability},
Phys.\ Rev.\ Lett.\  {\bf 108} (2012) 261604, 
{\tt [arXiv:1203.6180 [hep-th]]}.
%%CITATION = ARXIV:1203.6180;%%

I.~Kostov, {\em Three-point function of semiclassical states at weak coupling},
J.\ Phys.\ A {\bf 45} (2012) 494018, {\tt [arXiv:1205.4412 [hep-th]]}.
%%CITATION = ARXIV:1205.4412;%%

\bibitem{FW} O.~Foda and M.~Wheeler,
{\em Slavnov determinants, Yang-Mills structure constants, and discrete KP},
{\tt [arXiv:1203.5621 [hep-th]]}.
%%CITATION = ARXIV:1203.5621;%%

\bibitem{Serban} D.~Serban,
{\em A note on the eigenvectors of long-range spin chains and their scalar products},
JHEP {\bf 1301} (2013) 012, {\tt [[arXiv:1203.5842 [hep-th]]}.
%%CITATION = ARXIV:1203.5842;%%

\bibitem{FW2} O.~Foda and M.~Wheeler, 
{\em Partial domain wall partition functions},
JHEP {\bf 1207} (2012) 186, {\tt [arXiv:1205.4400 [math-ph]]}.
%%CITATION = ARXIV:1205.4400;%%

\bibitem{GromovVieira} 
N.~Gromov and P.~Vieira,
{\em Tailoring Three-Point Functions and Integrability IV. Theta-morphism}, 
JHEP \textbf{04} (2014), 068, 
{\tt [arXiv:1205.5288 [hep-th]]}.
%%CITATION = ARXIV:1205.5288;%%

\bibitem{KostovMatsuo} I.~Kostov and Y.~Matsuo,
{\em Inner products of Bethe states as partial domain wall partition functions},
JHEP {\bf 1210} (2012) 168, 
{\tt [arXiv:1207.2562 [hep-th]]}.
%%CITATION = ARXIV:1207.2562;%%

\bibitem{FW3} O.~Foda and M.~Wheeler,
{\em Variations on Slavnov's scalar product},
JHEP {\bf 1210} (2012) 096, 
{\tt [arXiv:1207.6871 [math-ph]]}.
%%CITATION = ARXIV:1207.6871;%%

\bibitem{Bissi} A.~Bissi, G.~Grignani and A.~V.~Zayakin,
{\em The $SO(6)$ Scalar Product and Three-Point Functions from Integrability},
{\tt [arXiv:1208.0100 [hep-th]]}.
%%CITATION = ARXIV:1208.0100;%%

\bibitem{Serbanfinite} D.~Serban,
{\em Eigenvectors and scalar products for long range interacting spin chains II: the finite size effects},
JHEP {\bf 1308} (2013) 128, {\tt [arXiv:1302.3350 [hep-th]]}.
%%CITATION = ARXIV:1302.3350;%%

\bibitem{FJKS} O.~Foda, Y.~Jiang, I.~Kostov and D.~Serban,
{\em A tree-level 3-point function in the $su(3)$-sector of planar ${\cal N}=4$ SYM},
JHEP {\bf 1310} (2013) 138, {\tt [arXiv:1302.3539 [hep-th]]}.
%%CITATION = ARXIV:1302.3539;%%

\bibitem{Kazama} Y.~Kazama, S.~Komatsu and T.~Nishimura,
{\em A new integral representation for the scalar products of Bethe states for the XXX spin chain},
JHEP {\bf 1309} (2013) 013, {\tt [arXiv:1304.5011 [hep-th]]}.
%%CITATION = ARXIV:1304.5011;%%

\bibitem{Wheeler} M.~Wheeler,
{\em Multiple integral formulae for the scalar product of on-shell and off-shell Bethe vectors in $SU(3)$-invariant models},
Nucl.\ Phys.\ B {\bf 875} (2013) 186, 
{\tt [arXiv:1306.0552 [math-ph]]}.
%%CITATION = ARXIV:1306.0552;%%

\bibitem{BK} E.~Bettelheim and I.~Kostov,
{\em Semi-classical analysis of the inner product of Bethe states},
J. Phys. A \textbf{47} (2014), 245401, 
{\tt [arXiv:1403.0358 [hep-th]]}.
%%CITATION = ARXIV:1403.0358;%%

\bibitem{MZ} J.~A.~Minahan and K.~Zarembo,
{\em The Bethe ansatz for ${\cal N}=4$ superYang-Mills},
JHEP {\bf 0303} (2003) 013, {\tt [arXiv:0212208 [hep-th]]}.
%%CITATION = HEP-TH/0212208;%%

\bibitem{BDS} N.~Beisert, V.~Dippel and M.~Staudacher,
{\em A Novel Long-Range Spin Chain and Planar $\mathcal{N}=4$ Super Yang-Mills},
JHEP \textbf{0407} (2004), 075, {\tt [arXiv:0405001 [hep-th]]}.
%%CITATION = ARXIV:0405001;%%

\bibitem{BKI} V. E. Korepin, N. M. Bogoliubov, A. G. Izergin,
{\em Quantum Inverse Scattering Method and Correlation Functions (Cambridge Monographs on Mathematical Physics},
Cambridge Monographs on Mathematical Physics, 1993.
%doi:10.1017/CBO9780511628832

\bibitem{Faddeev} F.~D.~Faddeev,
{\em How Algebraic Bethe Ansatz works for integrable model},
Les Houches lectures 1996, {\tt [arXiv:9605187 [hep-th]]}.
%%CITATION = ARXIV:9605187;%%

\bibitem{Nepomechie:1998jf}
R.~I.~Nepomechie,
{\em A Spin chain primer},
Int. J. Mod. Phys. B \textbf{13} (1999), 2973-2986, {\tt [aXiv:hep-th/9810032 [hep-th]]}.
%%CITATION = ARXIV:hep-th/9810032;%%

\bibitem{Gomez} C.~G\'omez, G.~Sierra and M.~Ruiz-Altaba,
{\em Quantum groups in two-dimensional physics},
Cambridge University Press (2011).
%doi:10.1017/CBO9780511628825

\bibitem{GaudinSP} M.~Gaudin,
{\em Bose Gas in One Dimension. I. The Closure Property of the Scattering Wavefunctions},
J. Math. Phys. \textbf{12} (1971), 1674.

\bibitem{GK} F.~G\"ohmann and V.~E.~Korepin,
{\em Solution of the quantum inverse problem},
J.~Phys.~A \textbf{33} (1999), 1199 {\tt [arXiv:9910253 [hep-th]]}.
%%CITATION = ARXIV:9910253;%%

\bibitem{Maillet} J.~M.~Maillet,
{\em Correlation functions of the XXZ Heisenberg spin chain: Bethe ansatz approach}, 
Proceedings of the International Congress of Mathematicians, Madrid, Spain, 2006.


\bibitem{Sklyanin:1980ij}
E.~K.~Sklyanin,
{\em Quantum version of the method of inverse scattering problem}, 
Zap. Nauchn. Semin. \textbf{95} (1980), 55-128 (Russian version), J. Soviet Math. \textbf{19} (1982), 1546–1596 (English version).

\bibitem{KM1} T.~Klose and T.~McLoughlin,
{\em Worldsheet Form Factors in AdS/CFT},
Phys.~Rev.~D 87 (2013), 026004 {\tt [arXiv:1208.2020 [hep-th]]}.
%%CITATION = ARXIV:1208.2020;%%

\bibitem{Mossel} J.~J.~Mossel,
{\em Quantum integrable models out of equilibrium} (2012), Ph. D. Thesis.

\bibitem{Watson} K.~M.~Watson,
{\em Some general relations between the photoproduction and scattering of pi mesons},
Phys.\ Rev.\  {\bf 95} (1954) 228.
%%CITATION = PHRVA,95,228;%%

\bibitem{Smirnov} F.~A.~Smirnov, 
{\em Form-factors in completely integrable models of quantum field theory},
Adv.\ Ser.\ Math.\ Phys.\  {\bf 14} (1992) 1.
%%CITATION = 00304,14,1;%%

\bibitem{KM2} T.~Klose and T.~McLoughlin,
{\em Comments on Worldsheet Form Factors in AdS/CFT}, 
J. Phys. A \textbf{47} (2014) no.5, 055401, 
{\tt [arXiv:1307.3506 [hep-th]]}.
%%CITATION = ARXIV:1307.3506;%%

\bibitem{Janik} R.~A.~Janik, 
{\em The $AdS_5 \times S^5$ superstring worldsheet S-matrix and crossing symmetry},
Phys.\ Rev.\ D {\bf 73} (2006) 086006, 
{\tt [arXiv:0603038 [hep-th]]}.
%%CITATION = HEP-TH/0603038;%%

\bibitem{BR} S.~Belliard and E.~Ragoucy,
{\em Nested Bethe ansatz for 'all' closed spin chains},
J.\ Phys.\ A {\bf 41} (2008) 295202, {\tt [arXiv:0804.2822 [math-ph]]}.
%%CITATION = ARXIV:0804.2822;%%

\end{thebibliography}
\end{document}